\documentclass[aps,prb,reprint,superscriptaddress,floatfix,showpacs]{revtex4-1}
\usepackage{amsmath,amsfonts,amssymb}
\usepackage{graphicx}
\usepackage[english]{babel}

\usepackage[colorlinks=true,urlcolor=blue,citecolor=red]{hyperref}

\begin{document}

\author{Emmanuel Arras}
\affiliation{SP2M, INAC, CEA, 38054 Grenoble, France}

\author{Fr\'{e}d\'{e}ric Lan\c{c}on}
\affiliation{SP2M, INAC, CEA, 38054 Grenoble, France}

\author{Ivetta Slipukhina}
\affiliation{SP2M, INAC, CEA, 38054 Grenoble, France}

\author{{\'E}ric Prestat}
\affiliation{SP2M, INAC, CEA, 38054 Grenoble, France}
\affiliation{Laboratorium f\"ur Elektronenmikroskopie, Karlsruher Institut f\"ur Technologie (KIT), 76128 Karlsruhe, Germany}

\author{Mauro Rovezzi}
\affiliation{European Synchrotron Radiation Facility (ESRF), 6 Rue Jules Horowitz, 38043 Grenoble, France}

\author{Samuel Tardif}
\affiliation{SP2M, INAC, CEA, 38054 Grenoble, France}
\affiliation{Institut N\'eel, CNRS-Universit\'e Joseph Fourier, BP166, 38402 Grenoble, France}

\author{Andrey Titov}
\affiliation{SP2M, INAC, CEA, 38054 Grenoble, France}
\affiliation{Institut N\'eel, CNRS-Universit\'e Joseph Fourier, BP166, 38402 Grenoble, France}
\affiliation{A. M. Prokhorov General Physics Institute, Russian Academy of Sciences, 38 Vavilov street, Moscow 119991 Russia}

\author{Pascale Bayle-Guillemaud}
\affiliation{SP2M, INAC, CEA, 38054 Grenoble, France}

\author{Francesco d'Acapito}
\affiliation{CNR-IOM-OGG GILDA CRG c/o ESRF 6 Rue Jules Horowitz, 38043 Grenoble, France}

\author{Andr{\'e} Barski}
\affiliation{SP2M, INAC, CEA, 38054 Grenoble, France}

\author{Vincent Favre-Nicolin}
\affiliation{SP2M, INAC, CEA, 38054 Grenoble, France}
\affiliation{Universit\'e Joseph Fourier (UJF), Grenoble, France}

\author{Matthieu Jamet}
\affiliation{SP2M, INAC, CEA, 38054 Grenoble, France}

\author{Jo{\"e}l Cibert}
\affiliation{Institut N\'eel, CNRS-Universit\'e Joseph Fourier, BP166, 38402 Grenoble, France}

\author{Pascal Pochet}
\affiliation{SP2M, INAC, CEA, 38054 Grenoble, France}
\email{pascal.pochet@cea.fr}

\title{Interface-driven phase separation in multifunctional materials: the case of GeMn ferromagnetic semiconductor}

\date{\today}

\begin{abstract}
We use extensive first principle simulations to show the major role played by interfaces in the mechanism of phase separation observed in semiconductor multifunctional materials.
We make an analogy with the precipitation sequence observed in over-saturated AlCu alloys, and replace the Guinier-Preston zones in this new context.
A new class of materials, the $\alpha$ phases, is proposed to understand the formation of the coherent precipitates observed in the GeMn system.
The interplay between formation and interface energies is analyzed for these phases and for the structures usually considered in the literature.
The existence of the  $\alpha$ phases is assessed with both theoretical and experimental arguments.
\end{abstract}

% 71.15.Nc  Total energy and cohesive energy calculations
% 75.50.Pp  Magnetic semiconductors
% 61.46.-w  Structure of nanoscale materials
% Note : 31.15.A- (Ab initio calculations)
%        is in: 31. Electronic structure of atoms and molecules: theory

\maketitle

% Personal version number :
%\textcolor{blue}{\textbf{version}: \texttt{\mdseries v02-\myisodate\today{T}\currenttime}}

\section*{Introduction}
Multifunctional materials have drawn a lot of attention in recent years because of their intrinsic scientific interest and for the huge technological possibilities they offer.
Indeed these materials simultaneously exhibit two or more characteristics that are usually incompatible: ferroelectricity and ferromagnetism for multiferroics, uncorrelated  thermal and electrical conductivity for thermoelectrics, or semiconductor (SC) character and ferromagnetism.
Some homogeneous materials are available, however they usually exhibit weak properties or are limited to very low temperatures \cite{RMP_2010}.
In order to open a broader range of opportunities, heterogeneous materials have been proposed as an efficient way to couple strong responses at room temperature \cite{CEWENNAN_2008}.
Among the multifunctional materials, those based on semiconductors yield great promises, and similar enhancements are expected in nanostructured media for thermoelectricity \cite{MINGO_2009} and magnetic semiconductivity \cite{OHNO_1998}. Regarding this last issue, it has been shown that depending on the solubility limit and the growth temperature, magnetic impurities tend to make a solid solution and form dilute magnetic semiconductor (DMS), or gather in solute-rich precipitates via a phase separation process \cite{RMP_2010}. The formation of elongated nano-clusters observed in several systems such as Al(Cr)N (Ref.~\onlinecite{Gu_2005}) or (Ge,Mn)-nanocolumns \cite{JAMET_2006} is a way of getting the attractive heterogeneous materials discussed above.
Controlling the growth thanks to a detailed understanding of the phase separation mechanism, is a prerequisite for their efficient use to design new multifunctional materials.

As a new perspective, it is well known from physical metallurgy that phase separation into an over-saturated solid solution can occur via a three-step mechanism, for example in the AlCu system \cite{GP_DECOMP}: i) the solute atoms diffuse in the host matrix to form Guinier-Preston (GP) zones, which are solute-rich zones (also referred to as $\theta''$) that keep the host matrix atomic structure; ii) these zones transform to a secondary metastable phase ($\theta'$, isostructural to CaF$_2$) that is fully coherent with the host matrix, thus minimizing interface energy ; iii) a last transformation occurs with the formation of the stable Al$_2$Cu phase $\theta$ which is structurally incompatible with the host matrix. It is clear in this sequence that interfaces play a major role, especially in the stabilization of the metastable $\theta'$ phase.

In this article, we show that this interface-driven mechanism is also relevant in the case of magnetic semiconductors, but with specificities not taken into account up to now.
In order to get a deeper insight into the problem, we consider the GeMn system as a prototype, in which either dilute Mn atoms \cite{CHANGGAN_2008,TANAKA_2010}, Mn-rich nanocolumns~\cite{JAMET_2006,AHLERS_2006,Li_2007} (NC), or precipitates of the stable phase \cite{RMP_2010,WANG_2008} are observed depending on the growth conditions.

% \textcolor{blue}{
The AlCu precipitation sequence can be applied in the case of the GeMn system with the $\theta$ phase being the $Mn_5Ge_3$ compound while the $\theta'$ and $\theta''$ phases as revealed in the nanocolumns remain unknown.
This analysis is more complete than the simple spinodal decomposition scheme \cite{RMP_2010} that only focuses on the first stage of the former sequence leading to a coherent solute-rich solid solution (i.e. DMS $\rightarrow$ $\theta''$).
Still, the precipitation sequence needs to be slightly adapted.
On one hand, in the case of AlCu, one deals  with compact metallic phases with atomic structures and chemical behaviors basically all compatible in essence, leading to easy mixing, formation of solid solutions and natural interfaces. On the other hand, in the case of dilute semiconductors, the involved phases display non compact structures (diamond or wurtzite) for the host, and much more compact phases for precipitates, with very little structural and chemical compatibility to one another\cite{ARRAS_PRB_2010-1}, leading to strong chemical segregation and incoherent interfaces.
% }

The thorough study of two types of GeMn materials in bulk conditions, followed by the introduction of their interfaces with the semiconductor matrix, allows us to unravel not only the mechanisms behind this variety of precipitates, but also the atomic structure of the GeMn nanocolumns.
This yields a new understanding and potential control over these complex and very promising heterogeneous systems.

\section{Method}
\label{method}

We use the density functional theory (DFT) framework, within the projector augmented-wave approach  (PAW) and the generalized gradient approximation (GGA) for the exchange-correlation energy, as implemented in the \textsc{abinit} code \cite{ABINIT_2,ABINIT_16}.
The details are those used in Ref.~\onlinecite{ARRAS_PRB_2010-1}.
In particular, the pseudopotentials have been generated with the \textsc{atompaw} code\cite{ATOMPAW-1}.
They correspond to valence electron states 4$s$ and 4$p$ for Ge and 3$d$ and 4$s$ for Mn.
The $k$-point meshes and the cutoff energy ensure an error on the total DFT energy lower than 0.1~eV/atom.
% k-point mesh : 6x6x6 for the nTmS.

Since we want to compare the stability of phases with different values of the Mn concentration, $x_{\text{Mn}} = 1 - x_{\text{Ge}}$,
we calculate their formation energies per atom:
\begin{equation}
E_{\text{F}} = \epsilon - x_{\text{Mn}}  \mu_{\text{Mn}} - x_{\text{Ge}} \mu_{\text{Ge}}
\label{equ:Ef}
\end{equation}
where $\epsilon$ is the energy per atom found with our first-principle calculations, $\mu_{\text{Mn}}$ and $\mu_{\text{Ge}}$ are the chemical potential of Mn and Ge, respectively.
When manganese atoms are introduced in a germanium matrix, the ultimate system obtained after annealing is composed of Mn$_{11}$Ge$_{8}$ grains embedded in the diamond lattice of germanium.
Therefore, these crystals, Mn$_{11}$Ge$_{8}$ and Ge, are the two reference systems we have considered to determine $\mu_{\text{Mn}}$ and $\mu_{\text{Ge}}$.
When two phases $\gamma$ and $\delta$, with different compositions, are coexisting, the chemical potentials can be derived by considering the thermodynamic semi-grand canonical potential as described in Ref.~\onlinecite{ARRAS_PRB_2010-1}:
\begin{subequations}
\begin{eqnarray}{}
\mu_{\text{Ge}} &=& (x_{\text{Mn}}^{\delta} \epsilon^{\gamma} - x_{\text{Mn}}^{\gamma} \epsilon^{\delta})
/ (x_{\text{Mn}}^{\delta} - x_{\text{Mn}}^{\gamma})\\
\mu_{\text{Mn}} &=& (x_{\text{Ge}}^{\delta} \epsilon^{\gamma} - x_{\text{Ge}}^{\gamma} \epsilon^{\delta})
/ (x_{\text{Ge}}^{\delta} - x_{\text{Ge}}^{\gamma})
\end{eqnarray}
\end{subequations}
When $\gamma$ is the Ge diamond lattice and $\delta$ is the Mn$_{11}$Ge$_{8}$ compounds, we get:
\begin{subequations}
\begin{eqnarray}{}
\mu_{\text{Ge}} &=&  \epsilon^{\gamma}\\
\mu_{\text{Mn}} &=& (19 \epsilon^{\delta} - 8 \epsilon^{\gamma}) / 11
\label{equ:mu_2}
\end{eqnarray}
\end{subequations}
It can be checked that the formation energies [Eq.~(1)] of the Ge lattice and the Mn$_{11}$Ge$_{8}$ compound (or a mixture of them) are zero, since they are the chosen reservoirs of atoms in the semi-grand canonical ensemble.
Note that a less natural, but reasonable, choice would have been to choose the compound Mn$_{5}$Ge$_{3}$ as a reference, since it has a composition close to that of Mn$_{11}$Ge$_{8}$ and has a simpler unit cell. The difference on the formation energies of the compounds studied here would only have been of the order of 10~meV.

This methodology has proven to well reproduce both the properties and the stability of all known (Ge, Mn) compounds \cite{ARRAS_PRB_2010-1}, despite the necessary and usual approximations used.

%%%%%%%%%%%%%%%%%%%%%%%%%%%%%
\begin{figure*}
\includegraphics[width=17.2cm]{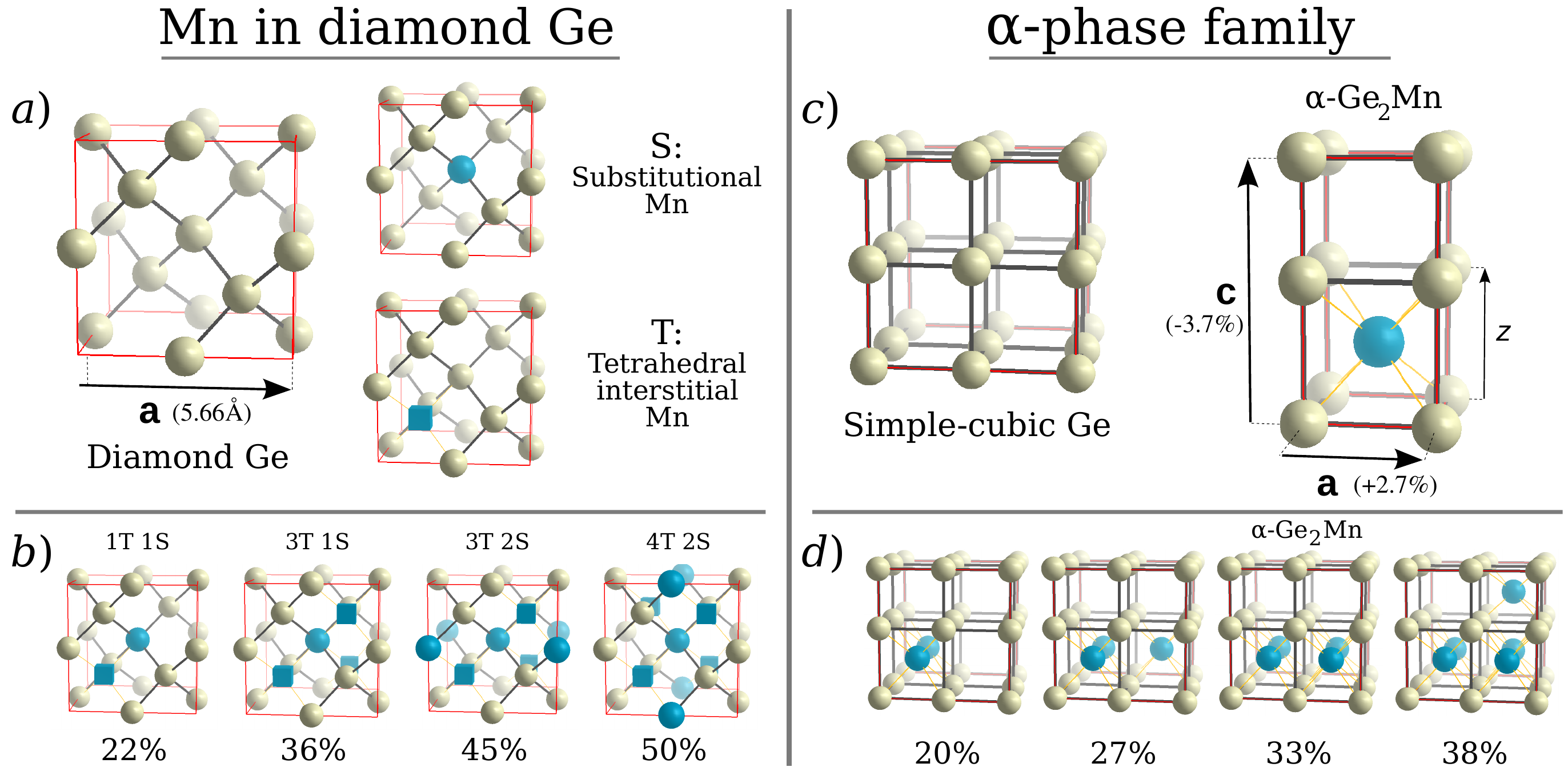}
\caption{(Color online)
The two structures tested: solid solution in the left column, intermetallic in the right column, with Mn contents $x_{\text{Mn}}$ indicated at the bottom.
\textit{a}) The diamond Ge lattice and the two most stable isolated Mn defects (S and T).
\textit{b}) Four typical combinations of the S and T defects.
\textit{c}) The simple-cubic Ge lattice and the $\alpha$ Ge$_2$Mn phase, isostructural to $\alpha$ FeSi$_2$.
\textit{d}) Four different combinations of Mn interstitials, representative of the $\alpha$-phase family.
For clarity reason, the displayed structures are unrelaxed.
}\label{fig_1}
\end{figure*}
%%%%%%%%%%%%%%%%%%%%%%%%%%%%%

\section{The solid-solution hypothesis}
\label{Sec:solid_solution}

We start with the assumption accepted so far \cite{JAMET_2006,DEVILLERS_2007,AHLERS_2006,Li_2007} that  the binding energy between magnetic impurities in the diamond structure drives the formation of a solid solution inside the nanocolumns \cite{FUKUSHIMA_2006}.
Those are thus supposed  to be domains where the concentration of Mn impurities is locally increased, yet with little effect on the diamond structure.
Since the Mn atoms could incorporate substitutionally or interstitially into the diamond Ge crystal lattice, we test the solid-solution assumption by considering a class of materials based on both Mn defects (see Fig.~\ref{fig_1}a).
More specifically, we use cubic cells of 5.658~\AA\ of side (8 atoms of Ge), and consider all the possible combinations of substitutional and tetrahedral interstitial Mn atoms up to a concentration of 60\%. In the following, we name $n$T$m$S a solid solution with $n$ Mn in tetrahedral interstitial sites and $m$ Mn in substitution per cubic unit cell, hence with a concentration $x_{\text{Mn}} = (n+m)/(n+8)$ where $n \leq 4$ and $m \leq 4$.
The restriction to these only two defects is based on observations by transmission electron microscopy showing a clear crystalline continuity between Mn-rich regions and diamond Ge along the [001] direction, thus discarding non crystal-aligned defects. In addition, our simulations of six different point defects involving one Mn atom
% \footnote{Stable lattice point defects associated with a Mn impurity: (S) Substitutional Mn atom replacing a germanium in the lattice ; (T) Mn in a Tetrahedral interstitial site; (H) Mn in an Hexagonal interstitial site; (V) split vacancy involving a substitutional Mn atom; (D) dumbbell interstitial with one lattice site shared by one Ge and one Mn; (FFCD) fourfold coordinated point defect \cite{Goedecker_2002} with one Mn atom involved.},
and of nine associated Mn dimers
% \footnote{
% \mbox{Mn} dimers in the Ge lattice are combination of the point defects listed in Ref.~\onlinecite{Note1}: S-S, T-T, S-T, S-H, S-V, T-V, H-V, D-D, and two Mn in one FFCD.},
confirm the energetic relevance of only the substitutional and tetrahedral interstitial defects (see appendix~\ref{annexe:point_defects}).
Finally, 27 non-equivalent configurations are tested (some examples are shown in Fig.~\ref{fig_1}b).
Ferromagnetic and antiferromagnetic orders have been considered, and the antiferromagnetic configurations turn out to have the lowest energies.

The evolution of the formation energy $E_{\text{F}}$ with the Mn content is reported on Fig.~\ref{fig_2}.
First, we can see a steep increase of $E_{\text{F}}$ at low concentration (the slope being 1.5~eV/Mn atom), in agreement with the low solubility of Mn in Ge \cite{SOLUB-GE_SIMU}.
Then at higher concentration, the interactions between defects lower the formation energy $E_{\text{F}}$.
Consequently, the curve differs from a straight line, and the formation energy reaches a maximum around 400~meV/atom, before decreasing.
The most significant result here is this hump-shaped curve with the maximum of the formation energy at 35\% of Mn, \textit{i.e.}, within the concentration range of 15-40\% for which the precipitates are observed experimentally \cite{JAMET_2006,DEVILLERS_2007,AHLERS_2006,Li_2007}.
The absence of a minimum of $E_{\text{F}}(x_{\text{Mn}})$ in this experimental range rules out a hypothetical stability of a diamond based coherent solid solution.

% \textcolor{blue}{
Accordingly, our calculations show that neither a spinodal decomposition, nor the analog of the first step of an AlCu precipitation sequence can occur in the bulk of the matrix.
In the latter case it would correspond to the so-called Guinier-Preston zones (also referred as $\theta''$ phase).
However, subsurface interstitials have been reported in Refs.~\onlinecite{ZHU_2004,CHANGGAN_2008}.
We propose that this interstitial-rich solid solution at the surface could be in fact the  $\theta''$ phase, forming  Guinier-Preston zones that will start the next precipitation step with a $\theta'$ phase inside nanocolumns.
In the following section, we propose a structure for such a $\theta'$ phase.
% }

\section{$\alpha$-phase structures}
Since the previous section has shown that GeMn solid-solution precipitates are not compatible with the $E_{\text{F}}(x_{\text{Mn}})$ curve, we need an equivalent of the $\theta'$ phase of AlCu, which is stabilized thanks to its natural interfaces with the host matrix. We propose a new class of materials based on the insertion of Mn into a simple-cubic crystal of Ge (Fig.~\ref{fig_1}c).

%%%%%%%%%%%%%%%%%%%%%%%%%%%%%
\begin{figure}

\includegraphics[width=8.6cm]{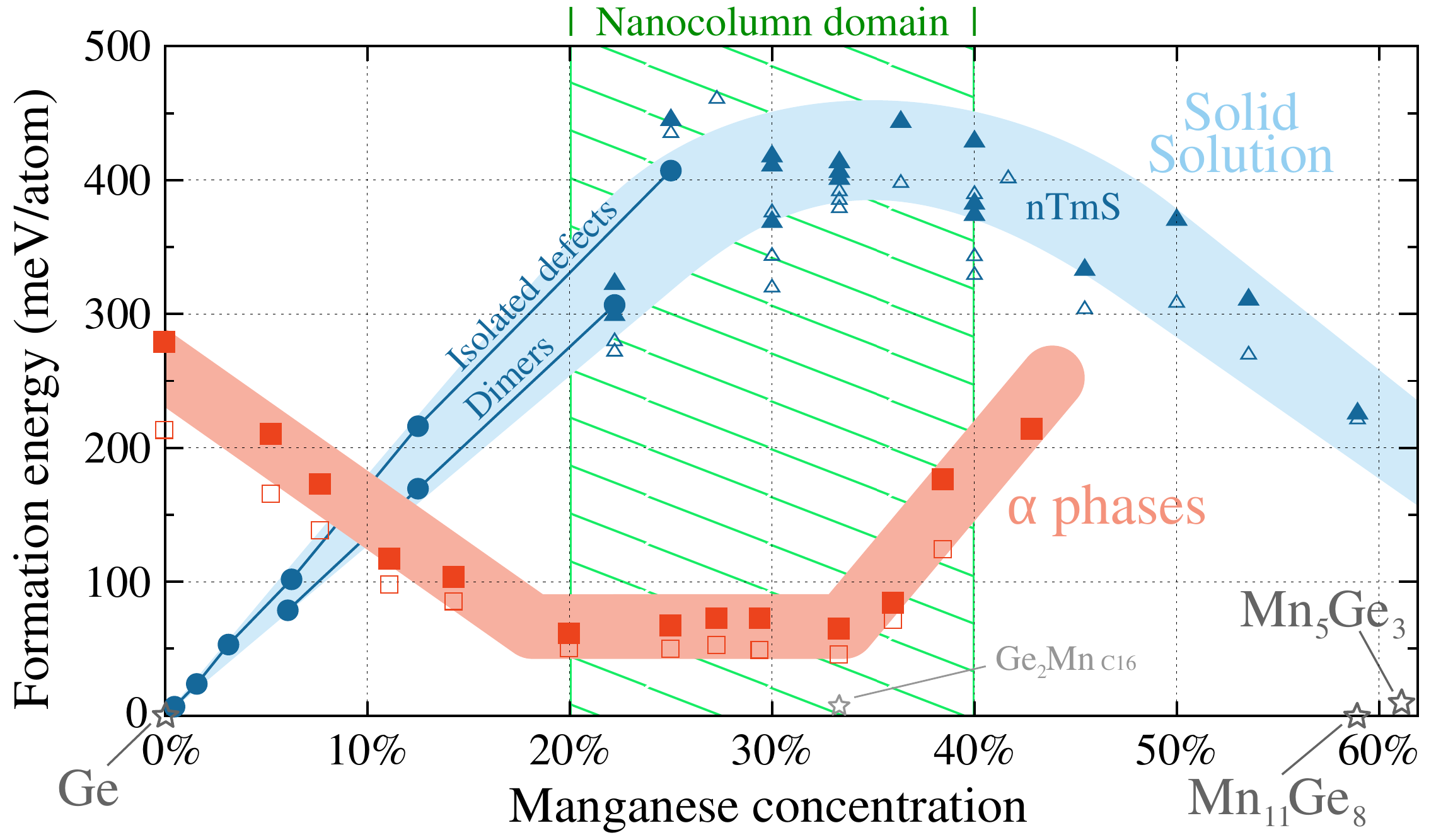}
\caption{(Color online)
Formation energy as a function of Mn content $x_{\text{Mn}}$, with diamond Ge and Ge$_8$Mn$_{11}$ being the references for the chemical potentials. Solid solution compounds are in blue (circles for low concentration DMSs, triangles for n\emph{T}m\emph{S} combinations), the $\alpha$ family in red (squares). Hollow markers denote fully-relaxed structures, \textit{i.e.}, both with respect to atomic coordinates and cell parameters.
Filled markers denote structures with the coordinates relaxed while the fixed cell parameters are commensurate with the experimental Ge lattice.
The dashed area (green online) shows the $x_{\text{Mn}}$-range of the experimentally observed nanocolumns.
The pastel broad lines are guides for the eyes for the solid-solution data on one hand, and the $\alpha$-family data on the other hand (respectively, blue and pink online).
}
\label{fig_2}

\end{figure}
%%%%%%%%%%%%%%%%%%%%%%%%%%%%%

Such a structure was proposed to describe MnGe$_4$, which is stable at high-pressure \cite{MnGe4_1990}. The exact structure was not determined experimentally, but -- from similarity with Hg$_4$Pt -- it was suggested to involve Mn atoms placed every four interstitial positions, which are at the center of the Ge cubes. It exhibits a striking structural proximity with diamond Ge, with a mismatch of only -2.5\% and -1.0\% along the $a$ and $c$ directions, respectively \cite{MnGe4_1990}. However, the calculated formation energy reaches 180~meV/atom \cite{ARRAS_PRB_2010-1}.

Applying the same approach as for the $n$T$m$S compounds, we have tested 45 different atomic configurations of Mn interstitials, using various cell sizes to have a complete energy diagram of all cubic Ge-based compounds. We have discovered a whole family of low energy structures, based on the Ge$_2$Mn phase shown in Fig.~\ref{fig_1}c, which is isostructural to $\alpha$ FeSi$_2$. We refer to it as the ``$\alpha$-phase family'' (see Fig.~\ref{fig_1}d). Such a variety in stoichiometry is common in the GeMn system, as illustrated by Mn$_3$Ge alloys \cite{ARRAS_PRB_2010-1}. The calculated equilibrium lattice parameters of the perfect $\alpha$ Ge$_2$Mn phase are $a$=$b$=2.90~\AA\ and $c$=5.45~\AA\ (Fig.~\ref{fig_1}c), leading to $c/a$=0.94. Thus, despite its simple-cubic crystal structure, this compound exhibits a lattice mismatch with bulk Ge lower than 4\% in all 3 directions, and a cell volume only 1.6\% larger, in agreement with observations of Ref.~\onlinecite{JAMET_2006}. Also, a decrease in the Mn content is accompanied by a regular decrease of the cell size, as expected from the interstitial character of Mn atoms, leading to a cell volume at 20\% Mn almost equal to that of bulk Ge, with $c/a=0.96$. 

Note that the new metastable Ge$_2$Mn compound C16 described in Ref.~\onlinecite{ARRAS_APL_2010} is directly related to the ``$\alpha$-phase family" but is distorted by a Jahn-Teller-type effect, making it strongly tetragonal and incompatible with the diamond structure.

Our calculations show that the $\alpha$ Ge$_2$Mn compound exhibits ferromagnetism, with a local moment on Mn atoms reaching 1~$\mu_{\text{B}}$, and a polarization P0=-80\% and P2=-83\% (with definitions of P0 and P2 as in Ref.~\onlinecite{MAZIN_1999}).

The formation energy of the $\alpha$ phases displays a broad minimum around 60~meV/atom, at Mn concentrations ranging from 20\% to 33\%. Above 33\%, Mn interstitial atoms must be added in the empty interstitial planes of $\alpha$ Ge$_2$Mn, at a formation energy cost of about 1~eV per Mn atom added. Below 20\%, Mn planes of $\alpha$ Ge$_2$Mn become less than half filled and the formation energy increases rapidly, until the unstable simple-cubic Ge is left.

Thus, contrary to the $n$T$m$S configurations, the $\alpha$-phases family gives rise to a second minimum of energy versus stoichiometry compatible with a segregation mechanism.
Furthermore, the flatness of the minimum in the 20 to 33\% range, agrees with the different Mn concentrations observed in the nanocolumns by Jamet \emph{et al.} (20\%--33\%) \cite{JAMET_2006,DEVILLERS_2007}, as well as in similar precipitates by Bougeard \emph{et al}. (15\%) \cite{AHLERS_2006} and by Li \emph{et al.} (17\%--30\%) \cite{Li_2007}.

Since the nanocolumns are embedded in the Ge matrix, the elastic energy has to be considered.
We do not calculate exactly this energy, which depends on the nanocolumn distribution, but an upper bound of it.
As shown in Fig.~\ref{fig_2}, we have calculated the formation energies of the possible structures either with their own lattices parameters, or constrained to be commensurate with the Ge lattice (hollow versus filled markers).
The last situation corresponds to a state in which all the elastic constraints are localized inside the columns. In the reality, the matrix would also adapt to the stress in order to globally minimize this energy.
Therefore, the energy difference  between a phase constrained or not constrained by the Ge diamond lattice is an upper bound of the total elastic energy.
These values ​​turn out to be quite small -- about 10~meV/atom -- compared to other energy terms, and are thus fairly accurate approximations of the values ​​of the elastic energy.
More precisely, the formation energy of a given compound, including the elastic energy due to its embedding in the matrix of germanium, is located between the hollow and filled markers in Fig.~\ref{fig_2}.

\section{Role of interfaces}
\label{section:interfaces}

The structural compatibility between diamond Ge and $\alpha$ phases is of major importance here. We now evaluate interface energies within the usual slab approximation, \textit{i.e.}, from simulations of alternating crystalline layers.
The surface energy $E_{\text{S}}$, between two phases $\gamma$ and $\delta$, is thus defined by
\begin{equation}
E_{\text{S}} = \frac{1}{S} \left(
E
- N^{\delta} E_{\text{F}}^{\delta}
- N^{\gamma} E_{\text{F}}^{\gamma}
\right)
\label{equ:Es}
\end{equation}
where $S$ is the total boundary area and $E$ is the total internal energy in the periodic cell, $N^{\delta}$ and $N^{\gamma}$ are the number of atoms in the layers of the respective phases, $E_{\text{F}}^{\delta}$ and $E_{\text{F}}^{\gamma}$ are the formation energies [Eq.~(\ref{equ:Ef})].
Appendix~\ref{annexe:surface_energy} details the relation between this Eq.~(\ref{equ:Es}) and the thermodynamic semi-grand canonical ensemble.
At equilibrium, the crystalline layers have not exactly the same lattice parameters parallel to the interface.
Thus, to minimize the strain imposed to the layers by the common computer cell periodicities, the size of this cell is relaxed perpendicularly to the interface during the calculation of $E$.
Accordingly, the formation energies $E_{\text{F}}^{\delta}$ and $E_{\text{F}}^{\gamma}$ in Eq.~(\ref{equ:Es}) are calculated with the same elastic conditions: fixed unit-cell periodicities except for the direction perpendicular to the interface.
The width of the layers is increased until we get a convergence for $E_{\text{S}}$ better than 2~meV/\AA$^2$. This condition is achieved when the width of each compound is greater than 15~\AA, \textit{i.e.}, a value that is of the same order of magnitude as the nanocolumn diameters.

The main interface orientations, [$1\,0\,0$] and [$1\,1\,0$], have been considered for the Ge/$\alpha$-Ge$_2$Mn boundary. For each of these orientations, all possible atomic arrangements at the interface have been tested, resulting in 22 different configurations.
The calculated interface energy between $\alpha$ Ge$_2$Mn and diamond Ge is found to be about 35~meV/\AA$^2$ for the lateral boundaries of the column (\textit{i.e.}, [$1\,0\,0$] and [$1\,1\,0$] orientations, the column axes being along [$0\,0\,1$]). For the interface perpendicular to the column axis, the interface energy is found larger: 63~meV/\AA$^2$, which is compatible with the cylindrical shape of the grains.
These energies are to be compared with the larger value of $E_{\text{S}}$ for the interfaces between Mn$_5$Ge$_3$ and diamond Ge (Ref.~\onlinecite{ARRAS_THESIS}): 80~meV/\AA$^2$ (Mn$_5$Ge$_3$ [$1\,1\,\bar{2}\,1$] on Ge [$1\,1\,0$] and ); 87~meV/\AA$^2$ (Mn$_5$Ge$_3$ [$0\,0\,0\,1$] on Ge [$0\,0\,1$]). Only the boundary corresponding to Mn$_5$Ge$_3$ [$0\,0\,0\,1$] on Ge [$1\,1\,1$] has a similar energy: 35~meV/\AA$^2$.

The transformation of the nanocolumns into rounded grains after annealing, shows that the nanocolumns are out of equilibrium.
Above, we have introduced and calculated the main terms of energy: chemical energy, elastic energy and interface energy.
We now want to take into account all of these terms and discuss the stability of the system depending on the particular distribution of Mn atoms.

First, we consider three types of large scale order:
i) full homogeneous dilution of Mn atoms in the whole Ge diamond matrix;
ii) nanocolumns with a high Mn concentration embedded in a matrix of germanium depleted in Mn;
and iii) spherical clusters under similar conditions.
The first case has been assumed in Refs.~\onlinecite{CHANGGAN_2008,TANAKA_2010}, with a growth at low temperature probably corresponding to a very limited atomic diffusion.
The second situation has been obtained at intermediate temperature \cite{JAMET_2006,AHLERS_2006}; the columnar shape has been linked to a predominance of surface diffusion during the growth \cite{FUKUSHIMA_2006}.
Finally, the spherical clusters have been obtained with a high temperature growth \cite{WANG_2008} or after annealing\cite{JAMET_2006}, and therefore involve a three-dimensional (3D) diffusion.

Second, we compute the total energy for different atomic local orders:
a) simple defects of the Ge diamond crystal;
b) $\alpha$-Ge$_2$Mn crystalline grains;
or c) Mn$_5$Ge$_3$ crystalline grains.
The details of the size and morphology are taken from Ref.~\onlinecite{DEVILLERS_2007}: average Mn concentration of 4.1\%, 30 000 columns per $\mu$m$^{2}$ (which together fix their mean diameter, 2~nm for $\alpha$ Ge$_2$Mn for instance), and spherical clusters of 10~nm in diameter.

%%%%%%%%%%%%%%%%%%%%%%%%%%%%%
\begin{figure}

\includegraphics[width=8.6cm]{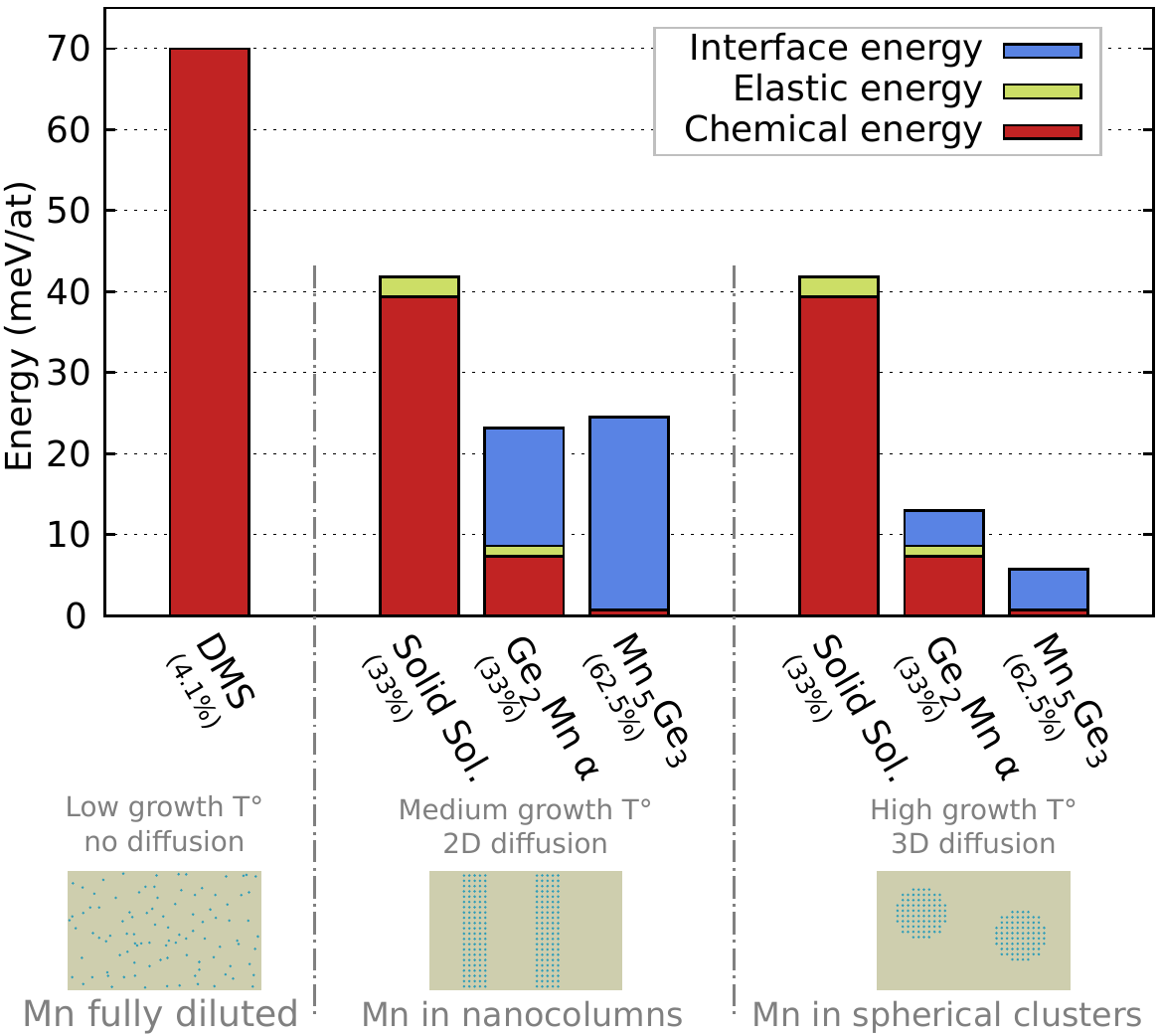}
\caption{(Color online)
Computed energies per atom of a Ge-Mn system, embedded in a Ge matrix depending on its crystalline structure: Ge-Mn solid solution, $\alpha$ Ge$_2$Mn, and $\eta$-Mn$_5$Ge$_3$.
The energy model includes chemical, elastic, and interface energies.
It is applied to three distinct Mn repartitions, which are metastable but experimentally observed depending on the growth conditions: i) Mn dilute in the host Ge matrix (DMS), ii) Mn atoms gathered inside nanocolumns, and iii) Mn inside spherical precipitates.
The assumed characteristics of the atomic diffusion in these three cases are indicated.
The experimental data describing these Mn repartitions are taken from Devillers \emph{et al.} \cite{DEVILLERS_2007}: global Mn concentration is 4.1\%, column density is 30000~$\mu$m$^{-2}$, diameter of spherical precipitates is 10~nm.
The zero of the energy corresponds to a phase separation between
diamond Ge and $\theta$-Mn11Ge8 (the atom reservoirs) with interfaces, free surfaces, and elastic constraints that are negligible.
}
\label{fig_3}

\end{figure}
%%%%%%%%%%%%%%%%%%%%%%%%%%%%%

Figure~\ref{fig_3} presents the energy contributions when the different chemical orders are considered.
First, it appears that the solid solution is ruled out by its very high total energy, in both phase-separation cases.
The case of full dilution has therefore to be limited to lower Mn contents.
The high surface-to-volume ratio of the nanocolumns ($S/V=$ 2~nm$^{-1}$ here as compared to 0.6~nm$^{-1}$ for a 10-nm sphere) is indeed not high enough to favor a columnar solid solution: the low chemical energy of $\alpha$ Ge$_2$Mn compensates for its interface energy.
On the other hand, $S/V$ is yet high enough to make the formation of Mn$_5$Ge$_3$ energetically unfavorable, because of its structural incompatibility with diamond-Ge, even if its formation energy is almost zero.
Indeed, the lateral coherence of $\alpha$ Ge$_2$Mn with the Ge crystal, and the resulting low-interface-energy compensate in this case for its higher formation energy.
Finally, in the case of spherical precipitates where the ratio $S/V$ is much smaller, the interface-energy advantage of $\alpha$ Ge$_2$Mn phase is not as crucial, and the Mn$_5$Ge$_3$ compound is found stable as observed experimentally.

% From these observations, a coherent scenario can be generalized for a three-stage precipitation process: first a solid solution exists until the limit of solubility is reached.
% Then, the high surface-diffusion during the sample growth\cite{DEVILLERS_2007} makes possible the precipitation of cylindrical grains of a phase coherent with the matrix.
% Eventually, the 3D diffusion leads to the stable compound with grains having a low surface-to-volume ratio, \textit{i.e.}, spheres in first approximation.

% \textcolor{blue}{
From these observations, we can now propose a coherent precipitation sequence in the line of the three-stage precipitation sequence of Al$_2$Cu (diluted $ \rightarrow \theta'' \rightarrow \theta' \rightarrow \theta $).
The first stage (DMS $\rightarrow \theta''$ ) corresponds to high surface-diffusion during the sample growth\cite{DEVILLERS_2007} and makes possible the  precipitation of Mn-rich domains in the subsurface of the sample.
Such a structure has not been studied in the present paper but is an extension of the reported subsurface interstitials \cite{ZHU_2004,CHANGGAN_2008}.
The second stage ($\theta'' \rightarrow \theta'$) corresponds to the germination of a coherent $\theta'$ phase from Mn interstitials. This phase has been identified as the proposed $\alpha$ phase, which is more likely to be stabilized in the form of nanocolumns (Fig.~\ref{fig_3}).
Finally, the last stage ($\theta' \rightarrow \theta$) is the transformation into the stable and less coherent Mn$_5$Ge$_3$ compound thanks to 3D diffusion.
This latter transformation also leads to a change of the shape of the precipitates from columns to round-shape precipitates.
% }

\section{Comparison to experimental results}

In previous sections, we have assessed the stability of the $\alpha$ phase based on theoretical arguments.
We now want to analyze our previous experimental characterizations of the GeMn nanocolumns \cite{JAMET_2006, DEVILLERS_2007, APL_ROVEZZI, TARDIF_APL_2010, TARDIF_PRB_2010} taking into account the structure that we propose with the $\alpha$ phase.
In addition, we first present new results of high-resolution electron microscopy.

\begin{figure}

\includegraphics[width=8.6cm]{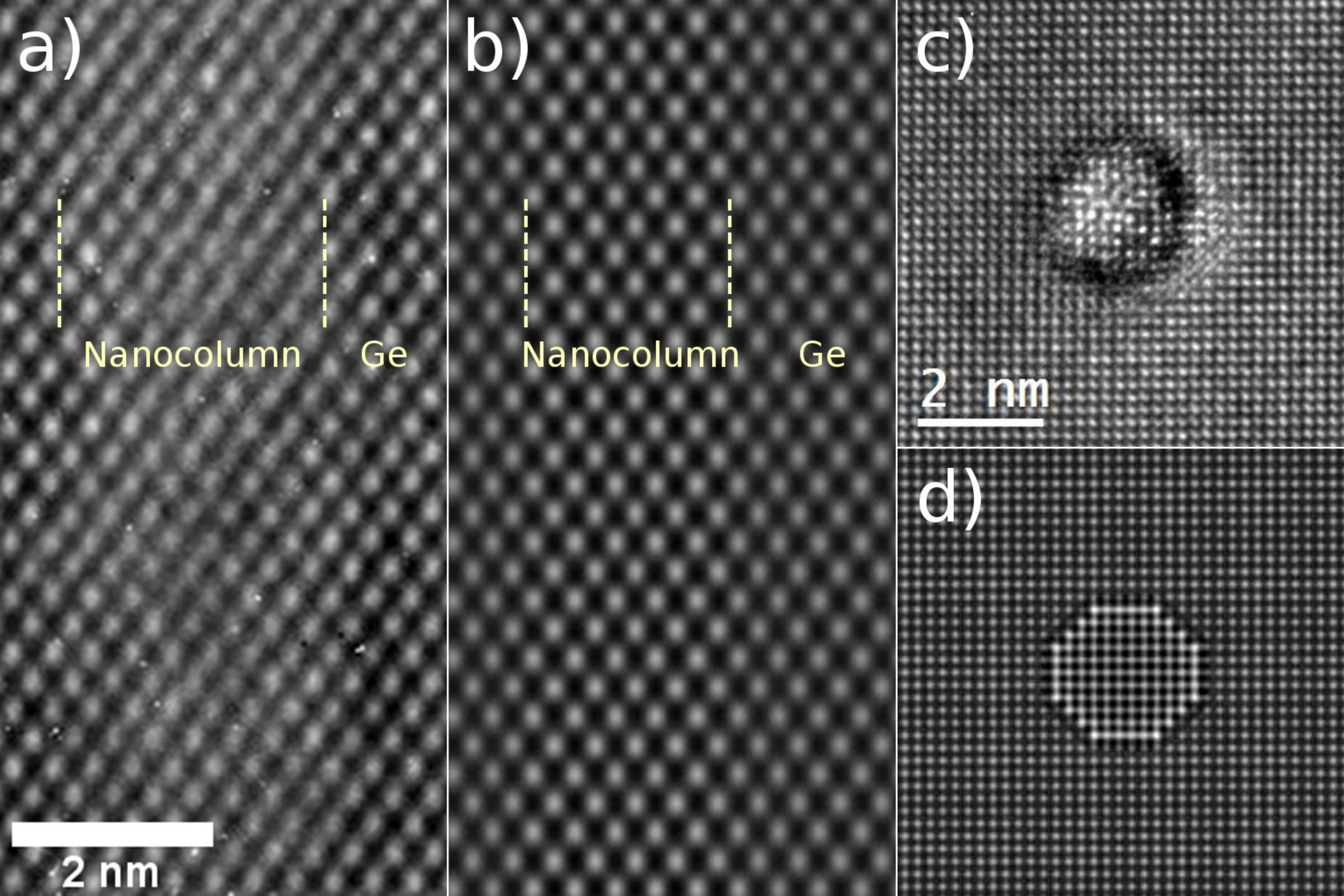}
\caption{Experimental (\textit{a} and \textit{c}) and simulated (\textit{b} and \textit{d}) TEM images of columns of about 2~nm in diameter. \textit{a} and \textit{b} show cross section along the $<$1 1 0$>$ direction, while \textit{c} and \textit{d} are plane views along the $<$0 0 1$>$ direction. Simulations were performed on an unrelaxed $\alpha$-phase nanocolumn using the \textsc{jems} software \cite{JEMS}.}
\label{fig_4}

\end{figure}

\paragraph{TEM ---}
Up to this work, the hypothesis made for the nanocolumn structure was the solid-solution phase in particular because of its strong similarity with the germanium matrix when observed by transmission electron microscopy (TEM) as shown in Figs.~\ref{fig_4}a and c.
Yet, the $<$0 0 1$>$ atomic rows of the $\alpha$ Ge$_2$Mn phase also form a square lattice compatible with the diamond Ge peaks positions in Fig.~\ref{fig_4}c.
The presence of two non-equivalent sub-lattices, one for Ge and one for Mn, corresponds to a super-structure which should be visible in plane view.
However, these sub-lattices generate two variants depending on the position of the cell origin with respect to the diamond-Ge cell, and these two variants are expected to coexist inside each column, because of the very low interface energy of the anti-phase boundaries bridging them (we calculate it to be lower than 10~meV/\AA$^2$). We have included these two variants to calculate a TEM image of an unrelaxed $\alpha$-Ge$_2$Mn nanocolumn in germanium (\textsc{jems} software \cite{JEMS}), in cross section (Fig.~\ref{fig_4}b) and in plane view (Fig.~\ref{fig_4}d). The simulated images show a striking resemblance between $\alpha$ phase and diamond Ge. The global contrast variation between the two phases is due to the manganese atoms, the germanium density being identical in both cases. Our calculations reproduce the dark rings around each column, corresponding to the structure discontinuity at the boundary.

\paragraph{Diffraction ---}
With the same inclusion of the two phase variants, we have computed the diffraction pattern of a collection of nanocolumns embedded in 100$\times$100 unit cells domains (see Ref.~\onlinecite{TARDIF_PRB_2010} for details on the column distribution and on the method). We find that the intensity is localized at the corresponding diamond diffraction peaks, and that the destructive interferences cancel the peaks due to the $\alpha$-phase sub-lattices.

%%%%%%%%%%%%%%%%%%%%%%%%%%%%%
%%%%%%%%%%%%%%%%%%%%%%%%%%%%%

%%%%%%%%%%%%%%%%%%%%%%%%%%%%%
% EXAFS figure
%%%%%%%%%%%%%%%%%%%%%%%%%%%%%
\begin{figure}
    \includegraphics[width=\columnwidth]{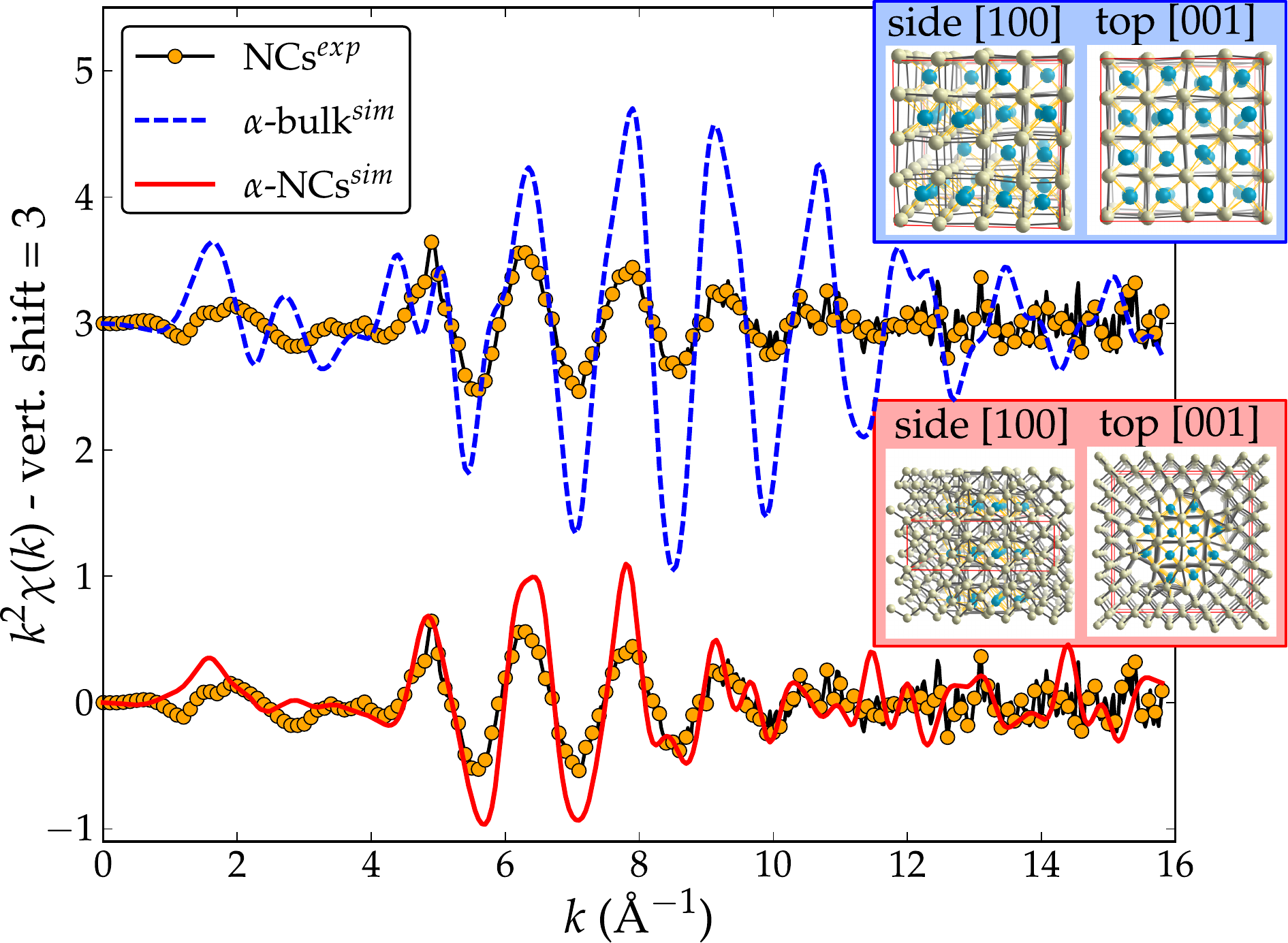}
    \caption{(Color online) Experimental data (Mn K-edge $k^2$-weighted EXAFS) for a sample with nanocolumns (NCs$^{exp}$) -- Ref.~\onlinecite{APL_ROVEZZI} -- compared with simulated spectra using the $\alpha$ Ge$_2$Mn phase in bulk form ($\alpha$-bulk$^{sim}$, top right panel) and as nanocolumns embedded in Ge ($\alpha$-NCs$^{sim}$, bottom right panel).}
\label{fig:exafs}
\end{figure}
%%%%%%%%%%%%%%%%%%%%%%%%%%%%%

\paragraph{EXAFS ---}
The Mn local atomic environment probed by extended X-ray absorption fine structure (EXAFS) spectroscopy at the Mn K-edge is also compatible with the proposed $\alpha$ Ge$_2$Mn phase when considered in the form of small nanocolumns (2~nm in diameter) embedded in a Ge matrix.
To demonstrate this, a different approach to the standard EXAFS analysis is adopted.
Instead of fitting the EXAFS signal with a set of theoretical scattering paths coming from Mn simple defects in Ge or clusters from well-known bulk structures (e.g., Mn$_5$Ge$_3$), as previously done by our group~\cite{APL_ROVEZZI} and others~\cite{Gunnella-2010}, we proceed in a qualitative analysis of the simulated EXAFS spectra based on molecular dynamics (MD) calculations of large relaxed cubic cells containing 96 atoms and 11.315~\AA\ of side, using the experimental lattice parameter of germanium~\cite{DAcapito:2011_SST}. Due to the fact that we are considering complex structures with multiple absorbing sites, this approach has the advantage to correctly account for both the configurational and dynamical disorders.
In fact, the EXAFS spectra are obtained by averaging the spectra of the single Mn atoms in the structure on the last 200 frames of the MD calculation%
~\footnote{The molecular dynamics calculations are carried out within the Density Functional Theory formalism. The canonical ensemble, NVT was used with 1000 time steps of 2~fs each with a target temperature of 300~K stabilized via a Nose thermostat.}
and successively averaging over all the Mn atoms in the cell.
For each configuration, a cluster of $R$~=~8~\AA\ around a central Mn atom is considered, and the simulations are carried out using the multiple scattering formalism with the Hedin-Lundqvist exchange potential as implemented in the \textsc{feff8} code~\cite{FEFF_8}.

In Fig.~\ref{fig:exafs} the simulated EXAFS spectra for the $\alpha$ Ge$_2$Mn phase in bulk form ($\alpha$-bulk$^{sim}$) and as nanocolumns embedded in Ge ($\alpha$-NCs$^{sim}$) are compared with the experimental data collected from a sample containing Mn-rich nanocolumns with a nominal Mn doping of 6\% -- details are given in Ref.~\onlinecite{APL_ROVEZZI}.
Both structures reproduce well the frequency of the experimental data, due to the value of the main Mn-Ge first shell distance of 2.51~\AA ( 2\% increase to respect bulk Ge) found by the previous calculations for the $\alpha$ phase.
There is a considerable difference in the amplitude of the oscillations and their shape in the low-$k$ region.
In fact, the double bump at 3~$<$~$k$~$<$~4 \AA$^{-1}$ that is the fingerprint of the nanocolumns~\cite{APL_ROVEZZI} and the peak at $k$~=~5~\AA$^{-1}$ are better reproduced when moving from the bulk form to embedded nano-objects.
This is demonstrated by the strong amplitude reduction when moving from the bulk ($\alpha$-bulk$^{sim}$) to the ideal nanocolumn ($\alpha$-NCs$^{sim}$).
This difference can be ascribed to the large number of Mn atoms located at the interface with the Ge matrix of which local environment is too highly distorted to generate a coherent oscillation. Indeed, for nanocolumns of 2~nm in diameter, half of the Mn atoms are at the interface and are displaced by more than 0.2~\AA\ as compared to perfect $\alpha$ Ge$_2$Mn. In fact, the first-shell coordination number $n_{\text{Mn-1}}$ found for $\alpha$-bulk$^{sim}$ is 8.5$\pm$0.5 while it is 5.5$\pm$0.5 for $\alpha$-NCs$^{sim}$. The strong decrease of $n_{\text{Mn-1}}$ illustrates a new way of interpreting the global coordination number of 4.0$\pm$0.5 found experimentally, not as a single low coordinated site but as a mixture of different Mn environments (in the nanocolumns, at their interfaces and diluted in Ge).

\paragraph{XAS, XMCD, and magnetometry ---}
Finally, the computed electronic properties of the $\alpha$ phases and their ferromagnetism are in line with the metallic character observed by X-ray absorption spectroscopy (XAS) at Mn L edge \cite{TARDIF_APL_2010} and the experimental magnetic local moment around 1~$\mu_{\text{B}}$ \cite{JAMET_2006,DEVILLERS_2007} from magnetometry and X-ray magnetic circular dichroism (XMCD).
In particular, it was shown that precise eigenvalues of the initial 2${p}$-states were necessary to obtain accurate DFT-calculated XAS and XMCD spectra~\cite{Tardif_GeMn_2011}.
The Curie temperature could not be calculated using standard techniques \cite{SLIPUKHINA_2009} because of the extreme variety of environments.
However the values of 340~K and 1.2~$\mu_{\text{B}}$/Mn measured for the high-pressure MnGe$_4$ compound \cite{MnGe4_1990} give an idea of the potentialities of these phases.

\section*{Conclusions}

We have considered two types of (Ge,Mn) compounds as possible candidates for coherent nanocolumn precipitates in germanium: i) a solid-solution model based on diamond Ge with substitutional and interstitial Mn atoms; ii) a new intermetallic compound with the Mn atoms at interstitial sites of a simple-cubic Ge crystal and referred to as  the $\alpha$ phase. Our results show that the first hypothesis of a  solid solution is not compatible with the experimental manganese concentration, which is between 15\% and 40\%. On the other hand, the proposed metastable $\alpha$ phase is coherent with the diamond structure of the matrix with which it has a low interfacial energy.

The solid solution has a negligible interfacial energy with diamond Ge, but a high formation energy.
Conversely, the compound Mn$_5$Ge$_3$ has a negligible formation energy, but a high interfacial energy.
The growth conditions yield columns, which have a high $S/V$ ratio.
We have shown that with such a geometry, the $\alpha$-phase nanocolumns have a favorable balance between these two energy terms.
Thus, the geometry imposed by the material growth stabilizes this otherwise metastable compound.

The $\alpha$ phase has a coherent interface with the germanium matrix and has a close connection with a simple cubic lattice.
Experimentally, it is difficult to distinguish it from an hypothetical diamond cubic Ge lattice with highly concentrated Mn impurities.
Although our argument in favor of the $\alpha$ phase as a constituent of the nanocolumns is based on theoretical results, we have shown these results are coherent with the experimental data available so far.

Finally, with the GeMn system as a prototype, we have adapted the three-step mechanism of precipitation in the AlCu system to rationalize the formation of coherent solute-rich precipitates in magnetic semi-conductors, the $\alpha$ phase playing the same role as the metastable Al$_2$Cu $\theta'$ phase.
Such an analysis would also prevail for others multifunctional semi-conductors where heterogeneous materials are used to tailor two or three different properties.

\begin{acknowledgments}
The simulations were performed at the CEA supercomputing center (CCRT), and part of this study was funded by ANR GEMO.
The EXAFS and DFT-MD simulations were carried out at the CORAL cluster of ESRF.
\end{acknowledgments}

%________________________________________________________________________________________
% Appendices should be speciﬁed using the \appendix % command which speciﬁes
%  that all following sections create with the \section commands are appendices.
%  If there is only one appendix, then the \appendix* command should be used instead.
\appendix

%________________________________________________________________________________________
\section{The manganese point defects in diamond germanium}
\label{annexe:point_defects}

\subsection{Introduction}
This appendix is the starting point of our analysis of the Ge--Mn solid solution done in section~\ref{Sec:solid_solution}. Indeed, we detail here the results regarding the point defects consisting of dilute Mn atoms in the Ge diamond structure and the dimers of such defects.
We address in particular the issue of their formation energy and their magnetic moment.

\begin{figure}

\includegraphics[width=8.6cm]{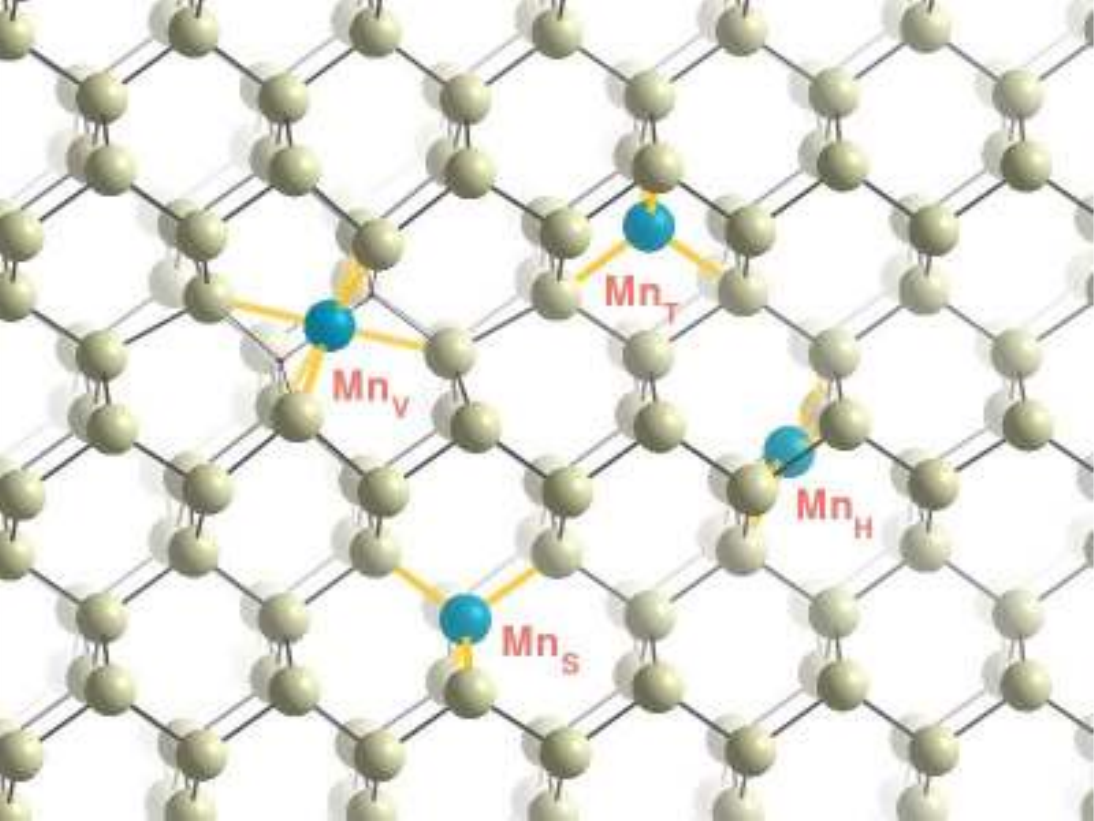}
\caption{(Color online) The Mn point defects in diamond Ge. Only (meta-)stable configurations are presented.}
\label{fig_6}

\end{figure}

We consider the six following defects:
\begin{itemize}
 \item Mn$_\text{S}$ : the substitutional position, which is the mere replacement of a Ge atom by a Mn atom;
 \item Mn$_\text{T}$ : the tetrahedral interstitial, in which the Mn atom is inserted in the bigger hole of the diamond structure;
 \item Mn$_\text{H}$ : the hexagonal interstitial, for which the Mn atom resides at the center of a Ge (non-planar) hexagon;
 \item Mn$_\text{V}$ : the split-vacancy, which is the final stable configuration of the initial combination of a vacancy and a neighboring substitutional Mn atom;
 \item Mn$_\text{D}$ : the ``dumbbell'' interstitial, for which one Mn and one Ge share one lattice site;
 \item Mn$_{\text{FFCD}}$ : the \textsc{ffcd} (\textit{fourfold coordinated point defect} \cite{Goedecker_2002}), which is a defect where bonds are rearranged keeping constant the coordination number of each atom. In our case, one of the two central atoms composing this defect is a Mn atom.
\end{itemize}
% From now on, the abbreviations will be used to refer to these defects.
We can split these defect into three categories, depending on the deviation $\Delta_N$ in the number of atoms compared to the pure diamond lattice: i) those with zero variation (Mn$_\text{S}$, Mn$_{\text{FFCD}}$), those for which one atom is removed (Mn$_\text{V}$), and those for which one atom is added (Mn$_\text{T}$, Mn$_\text{H}$, Mn$_\text{D}$). Among all these defects, two are found unstable upon structural relaxation, namely Mn$_\text{D}$ and Mn$_{\text{FFCD}}$. They become Mn$_\text{T}$ and Mn$_\text{S}$ respectively. The four others are presented on Fig.~\ref{fig_6}.

%%%%%%%%%%%%%%%%%%%%%%%%%%%%%
%%%%%%%%%%%%%%%%%%%%%%%%%%%%%

%/%/%/%/%/%/%/%/%/%/%/%/%/%/%/
\begin{table}[htbp]
\begin{minipage}{\linewidth}
\begin{center}
\begin{tabular}{|c||r|c|c|c|c|c|}
\hline
 Defect  & $\Delta_N$ & \multicolumn{3}{c|}{$E_{\text{F}}$ (eV)} &  $M_{\text{tot}}$  & $d_{\text{Mn-Ge}}$ \\
         &            & this work & Ref.~\onlinecite{SOLUB-GE_SIMU}\footnotemark[1] & Ref.~\onlinecite{CONTINENZA_2006-b}\footnotemark[1] & ($\mu_{\text{B}}$) &       (\AA)    \\
\hline
\hline
Mn$_\text{S}$   &     0      &      \textbf{1.5}   &  1.6   &  2.0  &    3    &  2.453 (-1.8\%)    \\
Mn$_\text{T}$   &    +1      &      2.0   &  2.3   &  3.0  &    5    &  2.586 (+3.6\%)    \\
Mn$_\text{H}$   &    +1      &      2.3   &        &       &    5    &  2.518 (+5.2\%)    \\
Mn$_\text{V}$   &    -1      &      2.8   &        &       &    4    &  2.761 (-12.1\%)   \\
\hline
\end{tabular}
\footnotetext[1]{The shown results have been adapted to take into account reference energy differences.}
\caption{Main characteristics of the four stable Mn simple defects in diamond Ge, simulated on Ge's equilibrium lattice parameter: formation energy $E_{\text{F}}$ (stabler defect is in bold), magnetic moment $M_{\text{tot}}$ and first neighbor distance $d_{\text{Mn-Ge}}$ (the deviation from pure diamond first neighbor distance is indicated). $\Delta_N$ is the change in atom number in the simulation cell, as compared to pure diamond.}\label{tab_1}
\end{center}
\end{minipage}
\end{table}
%/%/%/%/%/%/%/%/%/%/%/%/%/%/%/

%%%%%%%%%%%%%%%%%%%%%%%%%%%%%%%%%%
\begin{figure}[htbp]
\begin{center}
\includegraphics[width=8.5cm]{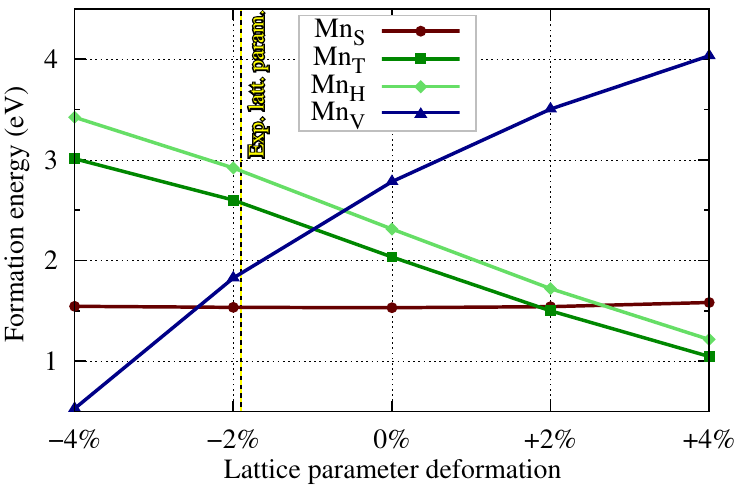}
\caption{Evolution of the formation energy of the 4 stable Mn simple defects versus isotropic strain of the diamond Ge. The experimental lattice parameter is indicated by the vertical dot dashed line at -1.9\%.}\label{fig_7}
\end{center}
\end{figure}
%%%%%%%%%%%%%%%%%%%%%%%%%%%%%%%%%%

%/%/%/%/%/%/%/%/%/%/%/%/%/%/%/
\begin{table}[htbp]
\begin{minipage}{\linewidth}
\begin{center}
\begin{tabular}{|l||r|c|c|c|c|}
\hline
 Defect  & $\Delta_N$ & $E_{\text{F}}$ &$\Delta_E^{\text{f-af}}$& $M_{\text{tot}}$ & $d_{\text{Mn-Mn}}$ \\
         &            &(eV/Mn)&    (eV)       & ($\mu_{\text{B}}$) &       (\AA)       \\
\hline
\hline
Mn$_{\text{2S}}$     &   0   &  1.7  &  -0.8  &      0     &  2.00 (-20.0\%)   \\
Mn$_{\text{2T}}$     &  +2   &  2.2  &   0.2  &      7     &  2.52 (+0.1\%)    \\
Mn$_{\text{ST}}$     &  +1   &  \textbf{1.3}  &  -0.3  &     0.3    &  2.45 (-2.0\%)    \\
Mn$_{\text{D2}}$     &  +1   &  2.4  &  -0.1  &      0     &  2.31 (-7.4\%)    \\
Mn$_{\text{FFCD2}}$  &  +0   &  1.9  &  -0.6  &      0     &  2.34 (-6.5\%)    \\
\hline
Mn$_{S}\footnotemark[1]$  & 0 & 1.7 & - & 3 & - \\
\hline
\end{tabular}
\footnotetext[1]{Simulated in a 32 atom box.}
\caption{Main characteristics of five Mn-dimers in diamond Ge, simulated on Ge's equilibrium lattice parameter.
The characteristic of the substitutional defect Mn$_{S}$, computed with the same simulation cell, is also given for comparison.
$\Delta_N$ is the change in atom number in the simulation cell, as compared to pure diamond.
$E_{\text{F}}$: minimum formation energy per Mn (stabler defect is in bold); $\Delta_E^{\text{f-af}}$: energy difference per Mn between ferromagnetic and antiferromagnetic configurations; $M_{\text{tot}}$: magnetic moment; $d_{\text{Mn-Mn}}$: Mn--Mn distance (the deviation from pure diamond first neighbor Ge distance is indicated too).
}\label{tab_2}
\end{center}
\end{minipage}
\end{table}
%/%/%/%/%/%/%/%/%/%/%/%/%/%/%/

\subsection{Computational details}
The DFT computation of heterogeneous defects in diamond Ge is not straightforward for three main reasons: i) due to the exchange-correlation approximation made for the calculation, the diamond-Ge gap is reduced to almost zero, and some localized electronic states of the defect could interact unrealistically with the diamond Ge conduction band;
ii) again due to the exchange-correlation approximation, the equilibrium lattice parameter of diamond Ge is off by almost 2\%, which is expected to have an impact on the defect characteristics;
iii) the computation of the formation energy of the  defects requires the use of reference energies, which is obvious for Ge, but not for Mn. We will now briefly detail how we intend to overcome these major issues.

\paragraph{Diamond-Ge gap and $k$-points ---}
To investigate the properties of an isolated defect, one must ensure that the defect is actually isolated in the simulations. However, the standard DFT methods rely on periodic boundary conditions to be efficient, yielding the simulation of not one, but an infinite array of defects.
The distance between image defects is thus a key parameter that controls the accuracy of the simulation. Here, we use cubic boxes that contain 64 Ge atom in the diamond structure.
In this case, the distance between image defects is about 11.5~\AA, \textit{i.e.}, large enough to prevent important interactions through the strain of the matrix. However, as noted earlier, electronic interactions may occur via the conduction band of the diamond Ge, artificially lowered by the exchange-correlation approximation.
A way to limit this phenomenon is to purposely use a coarse $k$-point mesh, along with special $k$-points\cite{KPT-SAMPLE}, to minimize the dispersive characteristic of the defect states.
Sufficient $k$-point mesh density must however be used to reproduce accurately the diamond Ge crystal.
In the following calculations, we use a $k$-point mesh that ensures errors lower than 5~meV/atom on the energy and 0.05\% on the lattice parameter of the Ge crystal.
It corresponds to a set of four $k$ points in the 64-atom box (by symmetry, only one inequivalent $k$-point), and for the two-atom diamond unit cell, to a $4 \times 4 \times 4$ Monkhorst-Pack grid \cite{MONKHORST-PACK} with the four usual shifts ([$\frac{1}{2}$ $\frac{1}{2}$ $\frac{1}{2}$], [$\frac{1}{2}$ $0$ $0$], [$0$ $\frac{1}{2}$ $0$], [$0$ $0$ $\frac{1}{2}$]).
This method is particularly efficient regarding the magnetic moment, as it gives the correct integer value in $\mu_{\text{B}}$ unit, \textit{i.e.}, the one obtained for bigger boxes, whereas a larger mesh, for example, $6 \times 6 \times 6$ in the 64-atom box, gives wrong fractional values.

\paragraph{Lattice parameter and strain ---}
The DFT calculated equilibrium lattice parameter of diamond Ge is found larger than the experimental value by 1.9\% with the usual Perdew-Burke-Ernzerhof (PBE) GGA approximation done in this article.
Hybrid functionals may reproduce the lattice parameter with a better agreement as shown in Ref.~\onlinecite{Stroppa_2011}.
Here, to directly evaluate the impact of the lattice parameter deviation on the characteristics of the defects, we compute them not only with this equilibrium lattice parameter, but also with 4 other lattice parameters shifted by: -4\% ($\sim$ +10~GPa); -2\% ($\sim$ +4~GPa); +2\% ($\sim$~-3~GPa); and +4\% ($\sim$~-5~GPa).

\paragraph{Formation energy ---}
The formation energy $E_{\text{F}}$ is given by a relation similar to Eq.~(\ref{equ:Ef}), where this time the energy is not a quantity by atom but for the defect:
\begin{equation}
E_{\text{F}} = E_{\text{tot}} - N_{\text{Mn}}  \mu_{\text{Mn}} - N_{\text{Ge}}  \mu_{\text{Ge}}
\label{equ:Ef_simple-def}
\end{equation}
where $E_{\text{tot}}$ is the DFT total energy of the cell, $N_{\text{Mn}}$ and $N_{\text{Ge}}$ are the partial numbers of atoms in the cell, and $\mu_{\text{Mn}}$ and $\mu_{\text{Ge}}$ the respective chemical potentials.
This formula is quite straightforward when one sees the chemical potentials as reference energy, as it has been done in many previous works.
The chemical potential of Ge, $\mu_{\text{Ge}}$, is easy to determine, namely, by considering the energy per atom of the pure Ge crystal.
However, $\mu_{\text{Mn}}$ has been given very different values in the previous works using many different references: the isolated Mn atom; the Mn atom as substitutional defect \cite{ZHU_2004}; the GeMn zinc-blend structure; the high temperature Mn $\gamma$ phase \cite{CONTINENZA_2006-b}; or the high pressure GeMn B20 phase \cite{SOLUB-GE_SIMU}.
Some authors do not even use references \cite{ZHU_2008,ZHU_2009}, but just give formation energy differences.
We have considered none of these methods and, as described in Sec.~\ref{method}, we choose diamond Ge and Mn$_{11}$Ge$_8$ as the reference atom reservoirs.

\subsection{Results}

\paragraph{Simple defects ---}

The results obtained for simple defects are gathered in Table~\ref{tab_1}. We can see that the substitutional Mn (Mn$_\text{S}$) is the most stable simple defect in diamond Ge, and exhibits a magnetic moment of 3~$\mu_B$, as already shown by previous works\cite{CONTINENZA_2006-b}. We find the value of 1.5~eV for its formation energy, which explains the very low solubility of Mn in diamond Ge, and the initial slope of $E_{\text{F}}(x_{\text{Mn}})$ in Fig.~\ref{fig_2}. Magnetic moments are in accordance with other theoretical investigations, when their simulation boxes are large enough, or $k$-point mesh coarse enough. Regarding the impact of the choice of lattice parameter on our results, the evolution of the formation energy versus the isotropic strain applied to the diamond is shown in Fig.~\ref{fig_7}.
As expected, we observe that the slope of the energy curve versus strain depends on the type of defect, namely, if there are more or less atoms than in pure diamond Ge. In the substitutional case Mn$_\text{S}$, in which the number of atoms is conserved, strain has almost no impact, and the formation energy given in Table~\ref{tab_1} is fairly accurate.
On the contrary, for interstitials (Mn$_\text{T}$ and Mn$_\text{H}$), a decrease of the lattice parameter (to come closer to the experimental lattice parameter) increases the formation energy. Oppositely, for the split vacancy Mn$_\text{V}$, the formation energy decreases with the lattice parameter. Thus, if Mn$_\text{S}$ appears to be definitively the most stable defect, and Mn$_\text{T}$ stabler than Mn$_\text{H}$, the question of the relative positioning of Mn$_\text{V}$ remains.

\paragraph{Dimers ---}
To go a step further in the study of these defects, we now study the several possible associations of two close Mn atoms, \textit{i.e.}, dimers, in particular the following ones:
\begin{itemize}
 \item Mn$_{\text{2S}}$ : two substitutional Mn at first neighbor sites (2 $\times$ Mn$_{\text{S}}$);
 \item Mn$_{\text{2T}}$ : two tetrahedral interstitial Mn (2 $\times$ Mn$_{\text{T}}$);
 \item Mn$_{\text{ST}}$ : one substitutional Mn and one tetrahedral interstitial Mn (Mn$_{\text{S}}$+Mn$_{\text{T}}$);
 \item Mn$_{\text{D2}}$ : the ``dumbbell'' interstitial, in which both atoms of the dimer are Mn, located on both sides of a lattice site;
 \item Mn$_{\text{FFCD2}}$ : the \textsc{ffcd}, in which both atoms composing the defect are Mn atoms.
\end{itemize}
Simulation box are here twice smaller than for simple defects, \textit{i.e.}, those of 32 atoms in the pure Ge crystal.
Each dimer was tested in both ferromagnetic and antiferromagnetic configurations.
Results are gathered in Table~\ref{tab_2}.
In addition, the dimers Mn$_{\text{V}}$+Mn$_{\text{T}}$ and Mn$_{\text{S}}$+Mn$_{\text{H}}$ are found to be unstable.
They relax toward Mn$_{\text{2S}}$ and Mn$_{\text{ST}}$ respectively.

Two points are worth noticing.
First, as already proposed \cite{CONTINENZA_2006-b}, the formation energy of Mn defects is reduced when they combine together, in particular from two Mn$_{\text{S}}$ into one Mn$_{\text{ST}}$.
Second, dimers with the lowest energy (below 2~eV/Mn) are antiferromagnetic or ferrimagnetic.

%________________________________________________________________________________________

\section{Interface energy}
\label{annexe:surface_energy}

We detail in this appendix the calculation of the interface energy as done in section~\ref{section:interfaces}.

In this article, we have considered the thermodynamic semi-grand canonical ensemble~\cite{ARRAS_PRB_2010-1} to calculate the formation energy per atom at $T$=0 of an alloy $\gamma$ of $A$ and $B$ atoms as
\begin{equation}
E_{\text{F}}^{\gamma} = \epsilon^{\gamma} - x_{A}^{\gamma}  \mu_{A} - x_{B}^{\gamma} \mu_{B}
\label{equ:Ef_A_B}
\end{equation}
where $\epsilon^{\gamma}$ is the DFT energy per atom, $\mu_{A}$ and $\mu_{B}$ are the chemical potentials of respectively $A$ and $B$ atoms. The quantity $x_{A}^{\gamma} = 1 - x_{B}^{\gamma}$ is the concentration of $A$ atoms.

In a similar way, we can compute the total formation energy $E_{\text{F}}$ of a given system of $N_A$ atoms $A$ and $N_B$ atoms $B$ distributed in two phases $\gamma$ and $\delta$:
\begin{equation}
E_{\text{F}} = E - N_{A}  \mu_{A} - N_{B}  \mu_{B}
\label{equ:Ef_with_interface}
\end{equation}
where the internal energy $E$ is the DFT total energy of the system.

Let $\mu^0_{A}$ and $\mu^0_{B}$ be the chemical potentials when the phases $\gamma$ and $\delta$ are at equilibrium (\textit{i.e.}, $\gamma$ and $\delta$ represent the atom reservoirs)~\cite{ARRAS_PRB_2010-1}:
\begin{subequations}
\label{equ:mu0}
\begin{eqnarray}{}
\mu^0_{A} &=& (x_{B}^{\delta} \epsilon^{\gamma} - x_{B}^{\gamma} \epsilon^{\delta})
/ (x_{B}^{\delta} - x_{B}^{\gamma})\\
\mu^0_{B} &=& (x_{A}^{\delta} \epsilon^{\gamma} - x_{A}^{\gamma} \epsilon^{\delta})
/ (x_{A}^{\delta} - x_{A}^{\gamma}).
\end{eqnarray}
\end{subequations}
Note that the chemical compositions of $\gamma$ and $\delta$ must be different for the denominators in Eq.(~\ref{equ:mu0}) to be nonzero.
We restrict our discussion to this case.
By definition, the formation energies of $\gamma$ and $\delta$ phases are null for $\mu_{A}=\mu^0_{A}$ and $\mu_{B}=\mu^0_{B}$ and then Eq.~(\ref{equ:Ef_A_B}) can also be written as
\begin{subequations}
\label{equ:Ef_mu0}
\begin{eqnarray}{}
E_{\text{F}}^{\gamma} = x_{A}^{\gamma} ( \mu^0_{A} - \mu_{A}) + x_{B}^{\gamma} ( \mu^0_{B} - \mu_{B} )\\
E_{\text{F}}^{\delta} = x_{A}^{\delta} ( \mu^0_{A} - \mu_{A}) + x_{B}^{\delta} ( \mu^0_{B} - \mu_{B} )
\end{eqnarray}
\end{subequations}

Now, we want to calculate the interface energy $E_{\text{I}}$ associated to the boundary between $\gamma$ and $\delta$.
Energy $E_{\text{I}}$ is defined as the energy difference between the formation energy $E_{\text{F}}$ of the combined system and the total energies of the separated phases in bulk forms:
\begin{equation}
E_{\text{I}} = E_{\text{F}}
- N^{\gamma} E_{\text{F}}^{\gamma}
- N^{\delta} E_{\text{F}}^{\delta}
\label{equ:Ef_interface}
\end{equation}
where $N^{\gamma}$ and $N^{\delta}$ are the respective number of atoms in $\gamma$ and $\delta$ phases.

This definition (\ref{equ:Ef_interface}), allows us to compare different metastable structures of the interface, for instance, a mere juxtaposition of both types of crystal cells or rather reconstructions of the interface zone with different local atomic concentrations.

The respective numbers $N^{\gamma}$ and $N^{\delta}$ of atoms in $\gamma$ and $\delta$ are related to the numbers $N_A$ and $N_B$ by
\begin{subequations}
\label{equ:numbers_A_B}
\begin{eqnarray}{}
N_{A} &=& x_{A}^{\gamma} N^{\gamma} + x_{A}^{\delta} N^{\delta}\\
N_{B} &=& x_{B}^{\gamma} N^{\gamma} + x_{B}^{\delta} N^{\delta}.
\end{eqnarray}
\end{subequations}
This system of equations defines the numbers $N^{\gamma}$ and $N^{\delta}$ if, and only if, the compositions of $\gamma$ and $\delta$ phases are different. This is indeed the case since $\gamma$ and $\delta$ correspond to the atom reservoirs that fix $\mu^0_{A}$ and $\mu^0_{B}$ [Eq.~(\ref{equ:mu0})].

From equations~(\ref{equ:Ef_mu0}) and~(\ref{equ:numbers_A_B}), we can relate the bulk part of the system energy to the deviations of the chemical potentials:
\begin{equation}
N^{\gamma} E_{\text{F}}^{\gamma} + N^{\delta} E_{\text{F}}^{\delta}
=
N_{A} ( \mu^0_{A} - \mu_{A}) + N_{B} ( \mu^0_{B} - \mu_{B} )
\label{equ:key_relation}
\end{equation}
and replace this bulk energy in Eq.(~\ref{equ:Ef_interface}) that becomes
\begin{equation}
E_{\text{I}} = E - N_{A}  \mu^0_{A} - N_{B}  \mu^0_{B} \quad \left(~\forall ( \mu_{A}, \mu_{B} )\,\right).
\label{equ:finale}
\end{equation}
Therefore, the interface energy $E_{\text{I}}$ of the boundary between the two phases is the formation energy $E^0_{\text{F}}$ of the system calculated for chemical potentials at which the phases are both at equilibrium so that their bulk formation energies vanish.

%________________________________________________________________________________________
%%% FIN annexe %%%

% \bibliography{biblio}

\begin{thebibliography}{0}%
\makeatletter
\providecommand \@ifxundefined [1]{%
 \@ifx{#1\undefined}
}%
\providecommand \@ifnum [1]{%
 \ifnum #1\expandafter \@firstoftwo
 \else \expandafter \@secondoftwo
 \fi
}%
\providecommand \@ifx [1]{%
 \ifx #1\expandafter \@firstoftwo
 \else \expandafter \@secondoftwo
 \fi
}%
\providecommand \natexlab [1]{#1}%
\providecommand \enquote  [1]{``#1''}%
\providecommand \bibnamefont  [1]{#1}%
\providecommand \bibfnamefont [1]{#1}%
\providecommand \citenamefont [1]{#1}%
\providecommand \href@noop [0]{\@secondoftwo}%
\providecommand \href [0]{\begingroup \@sanitize@url \@href}%
\providecommand \@href[1]{\@@startlink{#1}\@@href}%
\providecommand \@@href[1]{\endgroup#1\@@endlink}%
\providecommand \@sanitize@url [0]{\catcode `\\12\catcode `\$12\catcode
  `\&12\catcode `\#12\catcode `\^12\catcode `\_12\catcode `\%12\relax}%
\providecommand \@@startlink[1]{}%
\providecommand \@@endlink[0]{}%
\providecommand \url  [0]{\begingroup\@sanitize@url \@url }%
\providecommand \@url [1]{\endgroup\@href {#1}{\urlprefix }}%
\providecommand \urlprefix  [0]{URL }%
\providecommand \Eprint [0]{\href }%
\providecommand \doibase [0]{http://dx.doi.org/}%
\providecommand \selectlanguage [0]{\@gobble}%
\providecommand \bibinfo  [0]{\@secondoftwo}%
\providecommand \bibfield  [0]{\@secondoftwo}%
\providecommand \translation [1]{[#1]}%
\providecommand \BibitemOpen [0]{}%
\providecommand \bibitemStop [0]{}%
\providecommand \bibitemNoStop [0]{.\EOS\space}%
\providecommand \EOS [0]{\spacefactor3000\relax}%
\providecommand \BibitemShut  [1]{\csname bibitem#1\endcsname}%
\let\auto@bib@innerbib\@empty
%</preamble>
\end{thebibliography}%


\begin{thebibliography}{41}%
\makeatletter
\providecommand \@ifxundefined [1]{%
 \@ifx{#1\undefined}
}%
\providecommand \@ifnum [1]{%
 \ifnum #1\expandafter \@firstoftwo
 \else \expandafter \@secondoftwo
 \fi
}%
\providecommand \@ifx [1]{%
 \ifx #1\expandafter \@firstoftwo
 \else \expandafter \@secondoftwo
 \fi
}%
\providecommand \natexlab [1]{#1}%
\providecommand \enquote  [1]{``#1''}%
\providecommand \bibnamefont  [1]{#1}%
\providecommand \bibfnamefont [1]{#1}%
\providecommand \citenamefont [1]{#1}%
\providecommand \href@noop [0]{\@secondoftwo}%
\providecommand \href [0]{\begingroup \@sanitize@url \@href}%
\providecommand \@href[1]{\@@startlink{#1}\@@href}%
\providecommand \@@href[1]{\endgroup#1\@@endlink}%
\providecommand \@sanitize@url [0]{\catcode `\\12\catcode `\$12\catcode
  `\&12\catcode `\#12\catcode `\^12\catcode `\_12\catcode `\%12\relax}%
\providecommand \@@startlink[1]{}%
\providecommand \@@endlink[0]{}%
\providecommand \url  [0]{\begingroup\@sanitize@url \@url }%
\providecommand \@url [1]{\endgroup\@href {#1}{\urlprefix }}%
\providecommand \urlprefix  [0]{URL }%
\providecommand \Eprint [0]{\href }%
\providecommand \doibase [0]{http://dx.doi.org/}%
\providecommand \selectlanguage [0]{\@gobble}%
\providecommand \bibinfo  [0]{\@secondoftwo}%
\providecommand \bibfield  [0]{\@secondoftwo}%
\providecommand \translation [1]{[#1]}%
\providecommand \BibitemOpen [0]{}%
\providecommand \bibitemStop [0]{}%
\providecommand \bibitemNoStop [0]{.\EOS\space}%
\providecommand \EOS [0]{\spacefactor3000\relax}%
\providecommand \BibitemShut  [1]{\csname bibitem#1\endcsname}%
\let\auto@bib@innerbib\@empty
%</preamble>
\bibitem [{\citenamefont {Sato}\ \emph {et~al.}(2010)\citenamefont {Sato},
  \citenamefont {Bergqvist}, \citenamefont {Kudrnovsk\'y}, \citenamefont
  {Dederichs}, \citenamefont {Eriksson}, \citenamefont {Turek}, \citenamefont
  {Sanyal}, \citenamefont {Bouzerar}, \citenamefont {Katayama-Yoshida},
  \citenamefont {Dinh}, \citenamefont {Fukushima}, \citenamefont {Kizaki},\
  and\ \citenamefont {Zeller}}]{RMP_2010}%
  \BibitemOpen
  \bibfield  {author} {\bibinfo {author} {\bibfnamefont {K.}~\bibnamefont
  {Sato}}, \bibinfo {author} {\bibfnamefont {L.}~\bibnamefont {Bergqvist}},
  \bibinfo {author} {\bibfnamefont {J.}~\bibnamefont {Kudrnovsk\'y}}, \bibinfo
  {author} {\bibfnamefont {P.~H.}\ \bibnamefont {Dederichs}}, \bibinfo {author}
  {\bibfnamefont {O.}~\bibnamefont {Eriksson}}, \bibinfo {author}
  {\bibfnamefont {I.}~\bibnamefont {Turek}}, \bibinfo {author} {\bibfnamefont
  {B.}~\bibnamefont {Sanyal}}, \bibinfo {author} {\bibfnamefont
  {G.}~\bibnamefont {Bouzerar}}, \bibinfo {author} {\bibfnamefont
  {H.}~\bibnamefont {Katayama-Yoshida}}, \bibinfo {author} {\bibfnamefont
  {V.~A.}\ \bibnamefont {Dinh}}, \bibinfo {author} {\bibfnamefont
  {T.}~\bibnamefont {Fukushima}}, \bibinfo {author} {\bibfnamefont
  {H.}~\bibnamefont {Kizaki}}, \ and\ \bibinfo {author} {\bibfnamefont
  {R.}~\bibnamefont {Zeller}},\ }\href {\doibase 10.1103/RevModPhys.82.1633}
  {\bibfield  {journal} {\bibinfo  {journal} {Rev. Mod. Phys.}\ }\textbf
  {\bibinfo {volume} {82}},\ \bibinfo {pages} {1633} (\bibinfo {year}
  {2010})}\BibitemShut {NoStop}%
\bibitem [{\citenamefont {Nan}\ \emph {et~al.}(2008)\citenamefont {Nan},
  \citenamefont {Bichurin}, \citenamefont {Dong}, \citenamefont {Viehland},\
  and\ \citenamefont {Srinivasan}}]{CEWENNAN_2008}%
  \BibitemOpen
  \bibfield  {author} {\bibinfo {author} {\bibfnamefont {C.-W.}\ \bibnamefont
  {Nan}}, \bibinfo {author} {\bibfnamefont {M.~I.}\ \bibnamefont {Bichurin}},
  \bibinfo {author} {\bibfnamefont {S.}~\bibnamefont {Dong}}, \bibinfo {author}
  {\bibfnamefont {D.}~\bibnamefont {Viehland}}, \ and\ \bibinfo {author}
  {\bibfnamefont {G.}~\bibnamefont {Srinivasan}},\ }\href {\doibase
  10.1063/1.2836410} {\bibfield  {journal} {\bibinfo  {journal} {Journal of
  Applied Physics}\ }\textbf {\bibinfo {volume} {103}},\ \bibinfo {eid}
  {031101} (\bibinfo {year} {2008})}\BibitemShut {NoStop}%
\bibitem [{\citenamefont {Mingo}\ \emph {et~al.}(2009)\citenamefont {Mingo},
  \citenamefont {Hauser}, \citenamefont {Kobayashi}, \citenamefont
  {Plissonnier},\ and\ \citenamefont {Shakouri}}]{MINGO_2009}%
  \BibitemOpen
  \bibfield  {author} {\bibinfo {author} {\bibfnamefont {N.}~\bibnamefont
  {Mingo}}, \bibinfo {author} {\bibfnamefont {D.}~\bibnamefont {Hauser}},
  \bibinfo {author} {\bibfnamefont {N.~P.}\ \bibnamefont {Kobayashi}}, \bibinfo
  {author} {\bibfnamefont {M.}~\bibnamefont {Plissonnier}}, \ and\ \bibinfo
  {author} {\bibfnamefont {A.}~\bibnamefont {Shakouri}},\ }\href {\doibase
  10.1021/nl8031982} {\bibfield  {journal} {\bibinfo  {journal} {Nano Letters}\
  }\textbf {\bibinfo {volume} {9}},\ \bibinfo {pages} {711} (\bibinfo {year}
  {2009})}\BibitemShut {NoStop}%
\bibitem [{\citenamefont {Ohno}(1998)}]{OHNO_1998}%
  \BibitemOpen
  \bibfield  {author} {\bibinfo {author} {\bibfnamefont {H.}~\bibnamefont
  {Ohno}},\ }\href {\doibase 10.1126/science.281.5379.951} {\bibfield
  {journal} {\bibinfo  {journal} {Science}\ }\textbf {\bibinfo {volume}
  {281}},\ \bibinfo {pages} {951} (\bibinfo {year} {1998})}\BibitemShut
  {NoStop}%
\bibitem [{\citenamefont {Gu}\ \emph {et~al.}(2005)\citenamefont {Gu},
  \citenamefont {Wu}, \citenamefont {Liu}, \citenamefont {Singh}, \citenamefont
  {Newman},\ and\ \citenamefont {Smith}}]{Gu_2005}%
  \BibitemOpen
  \bibfield  {author} {\bibinfo {author} {\bibfnamefont {L.}~\bibnamefont
  {Gu}}, \bibinfo {author} {\bibfnamefont {S.~Y.}\ \bibnamefont {Wu}}, \bibinfo
  {author} {\bibfnamefont {H.}~\bibnamefont {Liu}}, \bibinfo {author}
  {\bibfnamefont {R.}~\bibnamefont {Singh}}, \bibinfo {author} {\bibfnamefont
  {N.}~\bibnamefont {Newman}}, \ and\ \bibinfo {author} {\bibfnamefont {D.~J.}\
  \bibnamefont {Smith}},\ }\href {\doibase 10.1016/j.jmmm.2004.11.446}
  {\bibfield  {journal} {\bibinfo  {journal} {Journal of Magnetism and Magnetic
  Materials}\ }\textbf {\bibinfo {volume} {290–291, Part 2}},\ \bibinfo
  {pages} {1395 } (\bibinfo {year} {2005})},\ \bibinfo {note} {proceedings of
  the Joint European Magnetic Symposia (JEMS' 04)}\BibitemShut {NoStop}%
\bibitem [{\citenamefont {Jamet}\ \emph {et~al.}(2006)\citenamefont {Jamet},
  \citenamefont {Barski}, \citenamefont {Devillers}, \citenamefont {Poydenot},
  \citenamefont {Dujardin}, \citenamefont {Bayle-Guillemaud}, \citenamefont
  {Rothman}, \citenamefont {Bellet-Amalric}, \citenamefont {Marty},
  \citenamefont {Cibert}, \citenamefont {Mattana},\ and\ \citenamefont
  {Tatarenko}}]{JAMET_2006}%
  \BibitemOpen
  \bibfield  {author} {\bibinfo {author} {\bibfnamefont {M.}~\bibnamefont
  {Jamet}}, \bibinfo {author} {\bibfnamefont {A.}~\bibnamefont {Barski}},
  \bibinfo {author} {\bibfnamefont {T.}~\bibnamefont {Devillers}}, \bibinfo
  {author} {\bibfnamefont {V.}~\bibnamefont {Poydenot}}, \bibinfo {author}
  {\bibfnamefont {R.}~\bibnamefont {Dujardin}}, \bibinfo {author}
  {\bibfnamefont {P.}~\bibnamefont {Bayle-Guillemaud}}, \bibinfo {author}
  {\bibfnamefont {J.}~\bibnamefont {Rothman}}, \bibinfo {author} {\bibfnamefont
  {E.}~\bibnamefont {Bellet-Amalric}}, \bibinfo {author} {\bibfnamefont
  {A.}~\bibnamefont {Marty}}, \bibinfo {author} {\bibfnamefont
  {J.}~\bibnamefont {Cibert}}, \bibinfo {author} {\bibfnamefont
  {R.}~\bibnamefont {Mattana}}, \ and\ \bibinfo {author} {\bibfnamefont
  {S.}~\bibnamefont {Tatarenko}},\ }\href {\doibase 10.1038/nmat1686}
  {\bibfield  {journal} {\bibinfo  {journal} {Nat. Mater.}\ }\textbf {\bibinfo
  {volume} {5}},\ \bibinfo {pages} {653} (\bibinfo {year} {2006})}\BibitemShut
  {NoStop}%
\bibitem [{\citenamefont {Murray}(1985)}]{GP_DECOMP}%
  \BibitemOpen
  \bibfield  {author} {\bibinfo {author} {\bibfnamefont {J.~L.}\ \bibnamefont
  {Murray}},\ }\href
  {http://www.ingentaconnect.com/content/maney/imtr/1985/00000030/00000001/art%
00011} {\bibfield  {journal} {\bibinfo  {journal} {International Metals
  Reviews}\ }\textbf {\bibinfo {volume} {30}},\ \bibinfo {pages} {211}
  (\bibinfo {year} {1985})}\BibitemShut {NoStop}%
\bibitem [{\citenamefont {Zeng}\ \emph {et~al.}(2008)\citenamefont {Zeng},
  \citenamefont {Zhang}, \citenamefont {van Benthem}, \citenamefont
  {Chisholm},\ and\ \citenamefont {Weitering}}]{CHANGGAN_2008}%
  \BibitemOpen
  \bibfield  {author} {\bibinfo {author} {\bibfnamefont {C.}~\bibnamefont
  {Zeng}}, \bibinfo {author} {\bibfnamefont {Z.}~\bibnamefont {Zhang}},
  \bibinfo {author} {\bibfnamefont {K.}~\bibnamefont {van Benthem}}, \bibinfo
  {author} {\bibfnamefont {M.~F.}\ \bibnamefont {Chisholm}}, \ and\ \bibinfo
  {author} {\bibfnamefont {H.~H.}\ \bibnamefont {Weitering}},\ }\href {\doibase
  10.1103/PhysRevLett.100.066101} {\bibfield  {journal} {\bibinfo  {journal}
  {Phys. Rev. Lett.}\ }\textbf {\bibinfo {volume} {100}},\ \bibinfo {eid}
  {066101} (\bibinfo {year} {2008})}\BibitemShut {NoStop}%
\bibitem [{\citenamefont {Yada}\ \emph {et~al.}(2010)\citenamefont {Yada},
  \citenamefont {Okazaki}, \citenamefont {Ohya},\ and\ \citenamefont
  {Tanaka}}]{TANAKA_2010}%
  \BibitemOpen
  \bibfield  {author} {\bibinfo {author} {\bibfnamefont {S.}~\bibnamefont
  {Yada}}, \bibinfo {author} {\bibfnamefont {R.}~\bibnamefont {Okazaki}},
  \bibinfo {author} {\bibfnamefont {S.}~\bibnamefont {Ohya}}, \ and\ \bibinfo
  {author} {\bibfnamefont {M.}~\bibnamefont {Tanaka}},\ }\href {\doibase
  10.1143/APEX.3.123002} {\bibfield  {journal} {\bibinfo  {journal} {App. Phys.
  Express}\ }\textbf {\bibinfo {volume} {3}},\ \bibinfo {pages} {123002}
  (\bibinfo {year} {2010})}\BibitemShut {NoStop}%
\bibitem [{\citenamefont {Bougeard}\ \emph {et~al.}(2006)\citenamefont
  {Bougeard}, \citenamefont {Ahlers}, \citenamefont {Trampert}, \citenamefont
  {Sircar},\ and\ \citenamefont {Abstreiter}}]{AHLERS_2006}%
  \BibitemOpen
  \bibfield  {author} {\bibinfo {author} {\bibfnamefont {D.}~\bibnamefont
  {Bougeard}}, \bibinfo {author} {\bibfnamefont {S.}~\bibnamefont {Ahlers}},
  \bibinfo {author} {\bibfnamefont {A.}~\bibnamefont {Trampert}}, \bibinfo
  {author} {\bibfnamefont {N.}~\bibnamefont {Sircar}}, \ and\ \bibinfo {author}
  {\bibfnamefont {G.}~\bibnamefont {Abstreiter}},\ }\href {\doibase
  10.1103/PhysRevLett.97.237202} {\bibfield  {journal} {\bibinfo  {journal}
  {Phys. Rev. Lett.}\ }\textbf {\bibinfo {volume} {97}},\ \bibinfo {eid}
  {237202} (\bibinfo {year} {2006})}\BibitemShut {NoStop}%
\bibitem [{\citenamefont {Li}\ \emph {et~al.}(2007)\citenamefont {Li},
  \citenamefont {Zeng}, \citenamefont {van Benthem}, \citenamefont {Chisholm},
  \citenamefont {Shen}, \citenamefont {{Nageswara Rao}}, \citenamefont {Dixit},
  \citenamefont {Feldman}, \citenamefont {Petukhov}, \citenamefont {Foygel},\
  and\ \citenamefont {Weitering}}]{Li_2007}%
  \BibitemOpen
  \bibfield  {author} {\bibinfo {author} {\bibfnamefont {A.~P.}\ \bibnamefont
  {Li}}, \bibinfo {author} {\bibfnamefont {C.}~\bibnamefont {Zeng}}, \bibinfo
  {author} {\bibfnamefont {K.}~\bibnamefont {van Benthem}}, \bibinfo {author}
  {\bibfnamefont {M.~F.}\ \bibnamefont {Chisholm}}, \bibinfo {author}
  {\bibfnamefont {J.}~\bibnamefont {Shen}}, \bibinfo {author} {\bibfnamefont
  {S.~V.~S.}\ \bibnamefont {{Nageswara Rao}}}, \bibinfo {author} {\bibfnamefont
  {S.~K.}\ \bibnamefont {Dixit}}, \bibinfo {author} {\bibfnamefont {L.~C.}\
  \bibnamefont {Feldman}}, \bibinfo {author} {\bibfnamefont {A.~G.}\
  \bibnamefont {Petukhov}}, \bibinfo {author} {\bibfnamefont {M.}~\bibnamefont
  {Foygel}}, \ and\ \bibinfo {author} {\bibfnamefont {H.~H.}\ \bibnamefont
  {Weitering}},\ }\href {\doibase 10.1103/PhysRevB.75.201201} {\bibfield
  {journal} {\bibinfo  {journal} {Phys. Rev. B}\ }\textbf {\bibinfo {volume}
  {75}},\ \bibinfo {eid} {201201} (\bibinfo {year} {2007})}\BibitemShut
  {NoStop}%
\bibitem [{\citenamefont {Wang}\ \emph {et~al.}(2008)\citenamefont {Wang},
  \citenamefont {Zou}, \citenamefont {Zhao}, \citenamefont {Han}, \citenamefont
  {Zhou},\ and\ \citenamefont {Wang}}]{WANG_2008}%
  \BibitemOpen
  \bibfield  {author} {\bibinfo {author} {\bibfnamefont {Y.}~\bibnamefont
  {Wang}}, \bibinfo {author} {\bibfnamefont {J.}~\bibnamefont {Zou}}, \bibinfo
  {author} {\bibfnamefont {Z.}~\bibnamefont {Zhao}}, \bibinfo {author}
  {\bibfnamefont {X.}~\bibnamefont {Han}}, \bibinfo {author} {\bibfnamefont
  {X.}~\bibnamefont {Zhou}}, \ and\ \bibinfo {author} {\bibfnamefont {K.~L.}\
  \bibnamefont {Wang}},\ }\href {\doibase 10.1063/1.2884527} {\bibfield
  {journal} {\bibinfo  {journal} {App. Phys. Lett.}\ }\textbf {\bibinfo
  {volume} {92}},\ \bibinfo {eid} {101913} (\bibinfo {year}
  {2008})}\BibitemShut {NoStop}%
\bibitem [{\citenamefont {Arras}\ \emph {et~al.}(2011)\citenamefont {Arras},
  \citenamefont {Caliste}, \citenamefont {Deutsch}, \citenamefont
  {Lan\c{c}on},\ and\ \citenamefont {Pochet}}]{ARRAS_PRB_2010-1}%
  \BibitemOpen
  \bibfield  {author} {\bibinfo {author} {\bibfnamefont {E.}~\bibnamefont
  {Arras}}, \bibinfo {author} {\bibfnamefont {D.}~\bibnamefont {Caliste}},
  \bibinfo {author} {\bibfnamefont {T.}~\bibnamefont {Deutsch}}, \bibinfo
  {author} {\bibfnamefont {F.}~\bibnamefont {Lan\c{c}on}}, \ and\ \bibinfo
  {author} {\bibfnamefont {P.}~\bibnamefont {Pochet}},\ }\href {\doibase
  10.1103/PhysRevB.83.174103} {\bibfield  {journal} {\bibinfo  {journal} {Phys.
  Rev. B}\ }\textbf {\bibinfo {volume} {83}},\ \bibinfo {pages} {174103}
  (\bibinfo {year} {2011})}\BibitemShut {NoStop}%
\bibitem [{\citenamefont {Gonze}\ \emph {et~al.}(2005)\citenamefont {Gonze},
  \citenamefont {Rignanese}, \citenamefont {Verstraete}, \citenamefont
  {Beuken}, \citenamefont {Pouillon}, \citenamefont {Caracas}, \citenamefont
  {Jollet}, \citenamefont {Torrent}, \citenamefont {Zerah}, \citenamefont
  {Mikami}, \citenamefont {Ghosez}, \citenamefont {Veithen}, \citenamefont
  {Raty}, \citenamefont {Olevano}, \citenamefont {Bruneval}, \citenamefont
  {Reining}, \citenamefont {Godby}, \citenamefont {Onida}, \citenamefont
  {Hamann}, \citenamefont {Allan}, \citenamefont {Zerah}, \citenamefont
  {Jollet}, \citenamefont {Torrent}, \citenamefont {Roy}, \citenamefont
  {Mikami}, \citenamefont {Ghosez}, \citenamefont {Raty},\ and\ \citenamefont
  {Allan}}]{ABINIT_2}%
  \BibitemOpen
  \bibfield  {author} {\bibinfo {author} {\bibfnamefont {X.}~\bibnamefont
  {Gonze}}, \bibinfo {author} {\bibfnamefont {G.-M.}\ \bibnamefont
  {Rignanese}}, \bibinfo {author} {\bibfnamefont {M.}~\bibnamefont
  {Verstraete}}, \bibinfo {author} {\bibfnamefont {J.-M.}\ \bibnamefont
  {Beuken}}, \bibinfo {author} {\bibfnamefont {Y.}~\bibnamefont {Pouillon}},
  \bibinfo {author} {\bibfnamefont {R.}~\bibnamefont {Caracas}}, \bibinfo
  {author} {\bibfnamefont {F.}~\bibnamefont {Jollet}}, \bibinfo {author}
  {\bibfnamefont {M.}~\bibnamefont {Torrent}}, \bibinfo {author} {\bibfnamefont
  {G.}~\bibnamefont {Zerah}}, \bibinfo {author} {\bibfnamefont
  {M.}~\bibnamefont {Mikami}}, \bibinfo {author} {\bibfnamefont
  {P.}~\bibnamefont {Ghosez}}, \bibinfo {author} {\bibfnamefont
  {M.}~\bibnamefont {Veithen}}, \bibinfo {author} {\bibfnamefont {J.-Y.}\
  \bibnamefont {Raty}}, \bibinfo {author} {\bibfnamefont {V.}~\bibnamefont
  {Olevano}}, \bibinfo {author} {\bibfnamefont {F.}~\bibnamefont {Bruneval}},
  \bibinfo {author} {\bibfnamefont {L.}~\bibnamefont {Reining}}, \bibinfo
  {author} {\bibfnamefont {R.}~\bibnamefont {Godby}}, \bibinfo {author}
  {\bibfnamefont {G.}~\bibnamefont {Onida}}, \bibinfo {author} {\bibfnamefont
  {D.}~\bibnamefont {Hamann}}, \bibinfo {author} {\bibfnamefont
  {D.}~\bibnamefont {Allan}}, \bibinfo {author} {\bibfnamefont
  {G.}~\bibnamefont {Zerah}}, \bibinfo {author} {\bibfnamefont
  {F.}~\bibnamefont {Jollet}}, \bibinfo {author} {\bibfnamefont
  {M.}~\bibnamefont {Torrent}}, \bibinfo {author} {\bibfnamefont
  {A.}~\bibnamefont {Roy}}, \bibinfo {author} {\bibfnamefont {M.}~\bibnamefont
  {Mikami}}, \bibinfo {author} {\bibfnamefont {P.}~\bibnamefont {Ghosez}},
  \bibinfo {author} {\bibfnamefont {J.-Y.}\ \bibnamefont {Raty}}, \ and\
  \bibinfo {author} {\bibfnamefont {D.}~\bibnamefont {Allan}},\ }\href@noop {}
  {\bibfield  {journal} {\bibinfo  {journal} {Zeit. Kristallogr.}\ }\textbf
  {\bibinfo {volume} {220}},\ \bibinfo {pages} {558} (\bibinfo {year}
  {2005})}\BibitemShut {NoStop}%
\bibitem [{\citenamefont {Torrent}\ \emph {et~al.}(2008)\citenamefont
  {Torrent}, \citenamefont {Jollet}, \citenamefont {Bottin}, \citenamefont
  {Zerah},\ and\ \citenamefont {Gonze}}]{ABINIT_16}%
  \BibitemOpen
  \bibfield  {author} {\bibinfo {author} {\bibfnamefont {M.}~\bibnamefont
  {Torrent}}, \bibinfo {author} {\bibfnamefont {F.}~\bibnamefont {Jollet}},
  \bibinfo {author} {\bibfnamefont {F.}~\bibnamefont {Bottin}}, \bibinfo
  {author} {\bibfnamefont {G.}~\bibnamefont {Zerah}}, \ and\ \bibinfo {author}
  {\bibfnamefont {X.}~\bibnamefont {Gonze}},\ }\href@noop {} {\bibfield
  {journal} {\bibinfo  {journal} {Comput. Mat. Science}\ }\textbf {\bibinfo
  {volume} {42}},\ \bibinfo {pages} {337} (\bibinfo {year} {2008})}\BibitemShut
  {NoStop}%
\bibitem [{\citenamefont {Holzwarth}\ \emph {et~al.}(2001)\citenamefont
  {Holzwarth}, \citenamefont {Tackett},\ and\ \citenamefont
  {Matthews}}]{ATOMPAW-1}%
  \BibitemOpen
  \bibfield  {author} {\bibinfo {author} {\bibfnamefont {N.}~\bibnamefont
  {Holzwarth}}, \bibinfo {author} {\bibfnamefont {A.}~\bibnamefont {Tackett}},
  \ and\ \bibinfo {author} {\bibfnamefont {G.}~\bibnamefont {Matthews}},\
  }\href@noop {} {\bibfield  {journal} {\bibinfo  {journal} {Computer Physics
  Communications}\ }\textbf {\bibinfo {volume} {135}},\ \bibinfo {pages}
  {329–347} (\bibinfo {year} {2001})}\BibitemShut {NoStop}%
\bibitem [{\citenamefont {Devillers}\ \emph {et~al.}(2007)\citenamefont
  {Devillers}, \citenamefont {Jamet}, \citenamefont {Barski}, \citenamefont
  {Poydenot}, \citenamefont {Bayle-Guillemaud}, \citenamefont {Bellet-Amalric},
  \citenamefont {Cherifi},\ and\ \citenamefont {Cibert}}]{DEVILLERS_2007}%
  \BibitemOpen
  \bibfield  {author} {\bibinfo {author} {\bibfnamefont {T.}~\bibnamefont
  {Devillers}}, \bibinfo {author} {\bibfnamefont {M.}~\bibnamefont {Jamet}},
  \bibinfo {author} {\bibfnamefont {A.}~\bibnamefont {Barski}}, \bibinfo
  {author} {\bibfnamefont {V.}~\bibnamefont {Poydenot}}, \bibinfo {author}
  {\bibfnamefont {P.}~\bibnamefont {Bayle-Guillemaud}}, \bibinfo {author}
  {\bibfnamefont {E.}~\bibnamefont {Bellet-Amalric}}, \bibinfo {author}
  {\bibfnamefont {S.}~\bibnamefont {Cherifi}}, \ and\ \bibinfo {author}
  {\bibfnamefont {J.}~\bibnamefont {Cibert}},\ }\href {\doibase
  10.1103/PhysRevB.76.205306} {\bibfield  {journal} {\bibinfo  {journal} {Phys.
  Rev. B}\ }\textbf {\bibinfo {volume} {76}},\ \bibinfo {eid} {205306}
  (\bibinfo {year} {2007})}\BibitemShut {NoStop}%
\bibitem [{\citenamefont {Fukushima}\ \emph {et~al.}(2006)\citenamefont
  {Fukushima}, \citenamefont {Sato}, \citenamefont {Katayama-Yoshida},\ and\
  \citenamefont {Dederichs}}]{FUKUSHIMA_2006}%
  \BibitemOpen
  \bibfield  {author} {\bibinfo {author} {\bibfnamefont {T.}~\bibnamefont
  {Fukushima}}, \bibinfo {author} {\bibfnamefont {K.}~\bibnamefont {Sato}},
  \bibinfo {author} {\bibfnamefont {H.}~\bibnamefont {Katayama-Yoshida}}, \
  and\ \bibinfo {author} {\bibfnamefont {P.~H.}\ \bibnamefont {Dederichs}},\
  }\href {\doibase 10.1143/JJAP.45.L416} {\bibfield  {journal} {\bibinfo
  {journal} {Japanese Journal of Applied Physics}\ }\textbf {\bibinfo {volume}
  {45}},\ \bibinfo {pages} {L416} (\bibinfo {year} {2006})}\BibitemShut
  {NoStop}%
\bibitem [{\citenamefont {Luo}\ \emph {et~al.}(2004)\citenamefont {Luo},
  \citenamefont {Zhang},\ and\ \citenamefont {Wei}}]{SOLUB-GE_SIMU}%
  \BibitemOpen
  \bibfield  {author} {\bibinfo {author} {\bibfnamefont {X.}~\bibnamefont
  {Luo}}, \bibinfo {author} {\bibfnamefont {S.~B.}\ \bibnamefont {Zhang}}, \
  and\ \bibinfo {author} {\bibfnamefont {S.-H.}\ \bibnamefont {Wei}},\ }\href
  {\doibase 10.1103/PhysRevB.70.033308} {\bibfield  {journal} {\bibinfo
  {journal} {Phys. Rev. B}\ }\textbf {\bibinfo {volume} {70}},\ \bibinfo
  {pages} {033308} (\bibinfo {year} {2004})}\BibitemShut {NoStop}%
\bibitem [{\citenamefont {Zhu}\ \emph {et~al.}(2004)\citenamefont {Zhu},
  \citenamefont {Weitering}, \citenamefont {Wang}, \citenamefont {Kaxiras},\
  and\ \citenamefont {Zhang}}]{ZHU_2004}%
  \BibitemOpen
  \bibfield  {author} {\bibinfo {author} {\bibfnamefont {W.}~\bibnamefont
  {Zhu}}, \bibinfo {author} {\bibfnamefont {H.~H.}\ \bibnamefont {Weitering}},
  \bibinfo {author} {\bibfnamefont {E.~G.}\ \bibnamefont {Wang}}, \bibinfo
  {author} {\bibfnamefont {E.}~\bibnamefont {Kaxiras}}, \ and\ \bibinfo
  {author} {\bibfnamefont {Z.}~\bibnamefont {Zhang}},\ }\href {\doibase
  10.1103/PhysRevLett.93.126102} {\bibfield  {journal} {\bibinfo  {journal}
  {Phys. Rev. Lett.}\ }\textbf {\bibinfo {volume} {93}},\ \bibinfo {pages}
  {126102} (\bibinfo {year} {2004})}\BibitemShut {NoStop}%
\bibitem [{\citenamefont {Takizawa}\ \emph {et~al.}(1990)\citenamefont
  {Takizawa}, \citenamefont {Sato}, \citenamefont {Endo},\ and\ \citenamefont
  {Shimada}}]{MnGe4_1990}%
  \BibitemOpen
  \bibfield  {author} {\bibinfo {author} {\bibfnamefont {H.}~\bibnamefont
  {Takizawa}}, \bibinfo {author} {\bibfnamefont {T.}~\bibnamefont {Sato}},
  \bibinfo {author} {\bibfnamefont {T.}~\bibnamefont {Endo}}, \ and\ \bibinfo
  {author} {\bibfnamefont {M.}~\bibnamefont {Shimada}},\ }\href {\doibase DOI:
  10.1016/0022-4596(90)90232-M} {\bibfield  {journal} {\bibinfo  {journal}
  {Journal of Solid State Chemistry}\ }\textbf {\bibinfo {volume} {88}},\
  \bibinfo {pages} {384 } (\bibinfo {year} {1990})}\BibitemShut {NoStop}%
\bibitem [{\citenamefont {Arras}\ \emph {et~al.}(2010)\citenamefont {Arras},
  \citenamefont {Slipukhina}, \citenamefont {Torrent}, \citenamefont {Caliste},
  \citenamefont {Deutsch},\ and\ \citenamefont {Pochet}}]{ARRAS_APL_2010}%
  \BibitemOpen
  \bibfield  {author} {\bibinfo {author} {\bibfnamefont {E.}~\bibnamefont
  {Arras}}, \bibinfo {author} {\bibfnamefont {I.}~\bibnamefont {Slipukhina}},
  \bibinfo {author} {\bibfnamefont {M.}~\bibnamefont {Torrent}}, \bibinfo
  {author} {\bibfnamefont {D.}~\bibnamefont {Caliste}}, \bibinfo {author}
  {\bibfnamefont {T.}~\bibnamefont {Deutsch}}, \ and\ \bibinfo {author}
  {\bibfnamefont {P.}~\bibnamefont {Pochet}},\ }\href {\doibase
  10.1063/1.3446837} {\bibfield  {journal} {\bibinfo  {journal} {Appl. Phys.
  Lett.}\ }\textbf {\bibinfo {volume} {96}},\ \bibinfo {eid} {231904} (\bibinfo
  {year} {2010})}\BibitemShut {NoStop}%
\bibitem [{\citenamefont {Mazin}(1999)}]{MAZIN_1999}%
  \BibitemOpen
  \bibfield  {author} {\bibinfo {author} {\bibfnamefont {I.~I.}\ \bibnamefont
  {Mazin}},\ }\href {\doibase 10.1103/PhysRevLett.83.1427} {\bibfield
  {journal} {\bibinfo  {journal} {Phys. Rev. Lett.}\ }\textbf {\bibinfo
  {volume} {83}},\ \bibinfo {pages} {1427} (\bibinfo {year}
  {1999})}\BibitemShut {NoStop}%
\bibitem [{\citenamefont {Arras}(2010)}]{ARRAS_THESIS}%
  \BibitemOpen
  \bibfield  {author} {\bibinfo {author} {\bibfnamefont {E.}~\bibnamefont
  {Arras}},\ }\emph {\bibinfo {title} {Theoretical study of the structure and
  stability of GeMn alloys in the context of spintronics (in French)}},\ \href
  {http://tel.archives-ouvertes.fr/tel-00489879/en/} {Ph.D. thesis},\ \bibinfo
  {school} {University of Grenoble - France} (\bibinfo {year}
  {2010})\BibitemShut {NoStop}%
\bibitem [{\citenamefont {Rovezzi}\ \emph {et~al.}(2008)\citenamefont
  {Rovezzi}, \citenamefont {Devillers}, \citenamefont {Arras}, \citenamefont
  {d'Acapito}, \citenamefont {Barski}, \citenamefont {Jamet},\ and\
  \citenamefont {Pochet}}]{APL_ROVEZZI}%
  \BibitemOpen
  \bibfield  {author} {\bibinfo {author} {\bibfnamefont {M.}~\bibnamefont
  {Rovezzi}}, \bibinfo {author} {\bibfnamefont {T.}~\bibnamefont {Devillers}},
  \bibinfo {author} {\bibfnamefont {E.}~\bibnamefont {Arras}}, \bibinfo
  {author} {\bibfnamefont {F.}~\bibnamefont {d'Acapito}}, \bibinfo {author}
  {\bibfnamefont {A.}~\bibnamefont {Barski}}, \bibinfo {author} {\bibfnamefont
  {M.}~\bibnamefont {Jamet}}, \ and\ \bibinfo {author} {\bibfnamefont
  {P.}~\bibnamefont {Pochet}},\ }\href {\doibase 10.1063/1.2949077} {\bibfield
  {journal} {\bibinfo  {journal} {Appl. Phys. Lett.}\ }\textbf {\bibinfo
  {volume} {92}},\ \bibinfo {eid} {242510} (\bibinfo {year}
  {2008})}\BibitemShut {NoStop}%
\bibitem [{\citenamefont {Tardif}\ \emph
  {et~al.}(2010{\natexlab{a}})\citenamefont {Tardif}, \citenamefont {Cherifi},
  \citenamefont {Jamet}, \citenamefont {Devillers}, \citenamefont {Barski},
  \citenamefont {Schmitz}, \citenamefont {Darowski}, \citenamefont {Thakur},
  \citenamefont {Cezar}, \citenamefont {Brookes}, \citenamefont {Mattana},\
  and\ \citenamefont {Cibert}}]{TARDIF_APL_2010}%
  \BibitemOpen
  \bibfield  {author} {\bibinfo {author} {\bibfnamefont {S.}~\bibnamefont
  {Tardif}}, \bibinfo {author} {\bibfnamefont {S.}~\bibnamefont {Cherifi}},
  \bibinfo {author} {\bibfnamefont {M.}~\bibnamefont {Jamet}}, \bibinfo
  {author} {\bibfnamefont {T.}~\bibnamefont {Devillers}}, \bibinfo {author}
  {\bibfnamefont {A.}~\bibnamefont {Barski}}, \bibinfo {author} {\bibfnamefont
  {D.}~\bibnamefont {Schmitz}}, \bibinfo {author} {\bibfnamefont
  {N.}~\bibnamefont {Darowski}}, \bibinfo {author} {\bibfnamefont
  {P.}~\bibnamefont {Thakur}}, \bibinfo {author} {\bibfnamefont {J.~C.}\
  \bibnamefont {Cezar}}, \bibinfo {author} {\bibfnamefont {N.~B.}\ \bibnamefont
  {Brookes}}, \bibinfo {author} {\bibfnamefont {R.}~\bibnamefont {Mattana}}, \
  and\ \bibinfo {author} {\bibfnamefont {J.}~\bibnamefont {Cibert}},\ }\href
  {\doibase 10.1063/1.3476343} {\bibfield  {journal} {\bibinfo  {journal} {App.
  Phys. Lett.}\ }\textbf {\bibinfo {volume} {97}},\ \bibinfo {eid} {062501}
  (\bibinfo {year} {2010}{\natexlab{a}})}\BibitemShut {NoStop}%
\bibitem [{\citenamefont {Tardif}\ \emph
  {et~al.}(2010{\natexlab{b}})\citenamefont {Tardif}, \citenamefont
  {Favre-Nicolin}, \citenamefont {Lan\c{c}on}, \citenamefont {Arras},
  \citenamefont {Jamet}, \citenamefont {Barski}, \citenamefont {Porret},
  \citenamefont {Bayle-Guillemaud}, \citenamefont {Pochet}, \citenamefont
  {Devillers},\ and\ \citenamefont {Rovezzi}}]{TARDIF_PRB_2010}%
  \BibitemOpen
  \bibfield  {author} {\bibinfo {author} {\bibfnamefont {S.}~\bibnamefont
  {Tardif}}, \bibinfo {author} {\bibfnamefont {V.}~\bibnamefont
  {Favre-Nicolin}}, \bibinfo {author} {\bibfnamefont {F.}~\bibnamefont
  {Lan\c{c}on}}, \bibinfo {author} {\bibfnamefont {E.}~\bibnamefont {Arras}},
  \bibinfo {author} {\bibfnamefont {M.}~\bibnamefont {Jamet}}, \bibinfo
  {author} {\bibfnamefont {A.}~\bibnamefont {Barski}}, \bibinfo {author}
  {\bibfnamefont {C.}~\bibnamefont {Porret}}, \bibinfo {author} {\bibfnamefont
  {P.}~\bibnamefont {Bayle-Guillemaud}}, \bibinfo {author} {\bibfnamefont
  {P.}~\bibnamefont {Pochet}}, \bibinfo {author} {\bibfnamefont
  {T.}~\bibnamefont {Devillers}}, \ and\ \bibinfo {author} {\bibfnamefont
  {M.}~\bibnamefont {Rovezzi}},\ }\href {\doibase 10.1103/PhysRevB.82.104101}
  {\bibfield  {journal} {\bibinfo  {journal} {Phys. Rev. B}\ }\textbf {\bibinfo
  {volume} {82}},\ \bibinfo {pages} {104101} (\bibinfo {year}
  {2010}{\natexlab{b}})}\BibitemShut {NoStop}%
\bibitem [{\citenamefont {Stadelmann}(2004)}]{JEMS}%
  \BibitemOpen
  \bibfield  {author} {\bibinfo {author} {\bibfnamefont {P.}~\bibnamefont
  {Stadelmann}},\ }\href
  {http://cimewww.epfl.ch/people/stadelmann/jemswebsite/jems.html} {\enquote
  {\bibinfo {title} {Jems, ems java version},}\ } (\bibinfo {year}
  {2004})\BibitemShut {NoStop}%
\bibitem [{\citenamefont {Gunnella}\ \emph {et~al.}(2010)\citenamefont
  {Gunnella}, \citenamefont {Morresi}, \citenamefont {Pinto}, \citenamefont
  {Cicco}, \citenamefont {Ottaviano}, \citenamefont {Passacantando},
  \citenamefont {Verna}, \citenamefont {Impellizzeri}, \citenamefont {Irrera},\
  and\ \citenamefont {d’Acapito}}]{Gunnella-2010}%
  \BibitemOpen
  \bibfield  {author} {\bibinfo {author} {\bibfnamefont {R.}~\bibnamefont
  {Gunnella}}, \bibinfo {author} {\bibfnamefont {L.}~\bibnamefont {Morresi}},
  \bibinfo {author} {\bibfnamefont {N.}~\bibnamefont {Pinto}}, \bibinfo
  {author} {\bibfnamefont {A.~D.}\ \bibnamefont {Cicco}}, \bibinfo {author}
  {\bibfnamefont {L.}~\bibnamefont {Ottaviano}}, \bibinfo {author}
  {\bibfnamefont {M.}~\bibnamefont {Passacantando}}, \bibinfo {author}
  {\bibfnamefont {A.}~\bibnamefont {Verna}}, \bibinfo {author} {\bibfnamefont
  {G.}~\bibnamefont {Impellizzeri}}, \bibinfo {author} {\bibfnamefont
  {A.}~\bibnamefont {Irrera}}, \ and\ \bibinfo {author} {\bibfnamefont
  {F.}~\bibnamefont {d’Acapito}},\ }\href
  {http://stacks.iop.org/0953-8984/22/i=21/a=216006} {\bibfield  {journal}
  {\bibinfo  {journal} {Journal of Physics: Condensed Matter}\ }\textbf
  {\bibinfo {volume} {22}},\ \bibinfo {pages} {216006} (\bibinfo {year}
  {2010})}\BibitemShut {NoStop}%
\bibitem [{\citenamefont {D’Acapito}(2011)}]{DAcapito:2011_SST}%
  \BibitemOpen
  \bibfield  {author} {\bibinfo {author} {\bibfnamefont {F.}~\bibnamefont
  {D’Acapito}},\ }\href {\doibase 10.1088/0268-1242/26/6/064004} {\bibfield
  {journal} {\bibinfo  {journal} {Semiconductor Science and Technology}\
  }\textbf {\bibinfo {volume} {26}},\ \bibinfo {pages} {064004} (\bibinfo
  {year} {2011})}\BibitemShut {NoStop}%
\bibitem [{Note1()}]{Note1}%
  \BibitemOpen
  \bibinfo {note} {The molecular dynamics calculations are carried out within
  the Density Functional Theory formalism. The canonical ensemble, NVT was used
  with 1000 time steps of 2~fs each with a target temperature of 300~K
  stabilized via a Nose thermostat.}\BibitemShut {Stop}%
\bibitem [{\citenamefont {Ankudinov}\ \emph {et~al.}(1998)\citenamefont
  {Ankudinov}, \citenamefont {Ravel}, \citenamefont {Rehr},\ and\ \citenamefont
  {Conradson}}]{FEFF_8}%
  \BibitemOpen
  \bibfield  {author} {\bibinfo {author} {\bibfnamefont {A.~L.}\ \bibnamefont
  {Ankudinov}}, \bibinfo {author} {\bibfnamefont {B.}~\bibnamefont {Ravel}},
  \bibinfo {author} {\bibfnamefont {J.~J.}\ \bibnamefont {Rehr}}, \ and\
  \bibinfo {author} {\bibfnamefont {S.~D.}\ \bibnamefont {Conradson}},\ }\href
  {\doibase 10.1103/PhysRevB.58.7565} {\bibfield  {journal} {\bibinfo
  {journal} {Phys. Rev. B}\ }\textbf {\bibinfo {volume} {58}},\ \bibinfo
  {pages} {7565} (\bibinfo {year} {1998})}\BibitemShut {NoStop}%
\bibitem [{\citenamefont {Tardif}(2011)}]{Tardif_GeMn_2011}%
  \BibitemOpen
  \bibfield  {author} {\bibinfo {author} {\bibfnamefont {S.}~\bibnamefont
  {Tardif}},\ }\emph {\bibinfo {title} {{GeMn} nanocolumns: magnetic and
  structural properties in light of synchrotron radiation}},\ \href
  {http://tel.archives-ouvertes.fr/tel-00585130/en/} {Ph.D. thesis},\ \bibinfo
  {school} {University of Grenoble, France} (\bibinfo {year}
  {2011})\BibitemShut {NoStop}%
\bibitem [{\citenamefont {Slipukhina}\ \emph {et~al.}(2009)\citenamefont
  {Slipukhina}, \citenamefont {Arras}, \citenamefont {Mavropoulos},\ and\
  \citenamefont {Pochet}}]{SLIPUKHINA_2009}%
  \BibitemOpen
  \bibfield  {author} {\bibinfo {author} {\bibfnamefont {I.}~\bibnamefont
  {Slipukhina}}, \bibinfo {author} {\bibfnamefont {E.}~\bibnamefont {Arras}},
  \bibinfo {author} {\bibfnamefont {P.}~\bibnamefont {Mavropoulos}}, \ and\
  \bibinfo {author} {\bibfnamefont {P.}~\bibnamefont {Pochet}},\ }\href@noop {}
  {\bibfield  {journal} {\bibinfo  {journal} {App. Phys. Lett.}\ }\textbf
  {\bibinfo {volume} {94}},\ \bibinfo {pages} {192505} (\bibinfo {year}
  {2009})}\BibitemShut {NoStop}%
\bibitem [{\citenamefont {Goedecker}\ \emph {et~al.}(2002)\citenamefont
  {Goedecker}, \citenamefont {Deutsch},\ and\ \citenamefont
  {Billard}}]{Goedecker_2002}%
  \BibitemOpen
  \bibfield  {author} {\bibinfo {author} {\bibfnamefont {S.}~\bibnamefont
  {Goedecker}}, \bibinfo {author} {\bibfnamefont {T.}~\bibnamefont {Deutsch}},
  \ and\ \bibinfo {author} {\bibfnamefont {L.}~\bibnamefont {Billard}},\ }\href
  {\doibase 10.1103/PhysRevLett.88.235501} {\bibfield  {journal} {\bibinfo
  {journal} {Phys. Rev. Lett.}\ }\textbf {\bibinfo {volume} {88}},\ \bibinfo
  {pages} {235501} (\bibinfo {year} {2002})}\BibitemShut {NoStop}%
\bibitem [{\citenamefont {Continenza}\ \emph {et~al.}(2006)\citenamefont
  {Continenza}, \citenamefont {Profeta},\ and\ \citenamefont
  {Picozzi}}]{CONTINENZA_2006-b}%
  \BibitemOpen
  \bibfield  {author} {\bibinfo {author} {\bibfnamefont {A.}~\bibnamefont
  {Continenza}}, \bibinfo {author} {\bibfnamefont {G.}~\bibnamefont {Profeta}},
  \ and\ \bibinfo {author} {\bibfnamefont {S.}~\bibnamefont {Picozzi}},\ }\href
  {\doibase 10.1103/PhysRevB.73.035212} {\bibfield  {journal} {\bibinfo
  {journal} {Phys. Rev. B}\ }\textbf {\bibinfo {volume} {73}},\ \bibinfo {eid}
  {035212} (\bibinfo {year} {2006})}\BibitemShut {NoStop}%
\bibitem [{\citenamefont {Makov}\ \emph {et~al.}(1996)\citenamefont {Makov},
  \citenamefont {Shah},\ and\ \citenamefont {Payne}}]{KPT-SAMPLE}%
  \BibitemOpen
  \bibfield  {author} {\bibinfo {author} {\bibfnamefont {G.}~\bibnamefont
  {Makov}}, \bibinfo {author} {\bibfnamefont {R.}~\bibnamefont {Shah}}, \ and\
  \bibinfo {author} {\bibfnamefont {M.~C.}\ \bibnamefont {Payne}},\ }\href
  {\doibase 10.1103/PhysRevB.53.15513} {\bibfield  {journal} {\bibinfo
  {journal} {Phys. Rev. B}\ }\textbf {\bibinfo {volume} {53}},\ \bibinfo
  {pages} {15513} (\bibinfo {year} {1996})}\BibitemShut {NoStop}%
\bibitem [{\citenamefont {Monkhorst}\ and\ \citenamefont
  {Pack}(1976)}]{MONKHORST-PACK}%
  \BibitemOpen
  \bibfield  {author} {\bibinfo {author} {\bibfnamefont {H.~J.}\ \bibnamefont
  {Monkhorst}}\ and\ \bibinfo {author} {\bibfnamefont {J.~D.}\ \bibnamefont
  {Pack}},\ }\href {\doibase 10.1103/PhysRevB.13.5188} {\bibfield  {journal}
  {\bibinfo  {journal} {Phys. Rev. B}\ }\textbf {\bibinfo {volume} {13}},\
  \bibinfo {pages} {5188} (\bibinfo {year} {1976})}\BibitemShut {NoStop}%
\bibitem [{\citenamefont {Stroppa}\ \emph {et~al.}(2011)\citenamefont
  {Stroppa}, \citenamefont {Kresse},\ and\ \citenamefont
  {Continenza}}]{Stroppa_2011}%
  \BibitemOpen
  \bibfield  {author} {\bibinfo {author} {\bibfnamefont {A.}~\bibnamefont
  {Stroppa}}, \bibinfo {author} {\bibfnamefont {G.}~\bibnamefont {Kresse}}, \
  and\ \bibinfo {author} {\bibfnamefont {A.}~\bibnamefont {Continenza}},\
  }\href {\doibase 10.1103/PhysRevB.83.085201} {\bibfield  {journal} {\bibinfo
  {journal} {Phys. Rev. B}\ }\textbf {\bibinfo {volume} {83}},\ \bibinfo
  {pages} {085201} (\bibinfo {year} {2011})}\BibitemShut {NoStop}%
\bibitem [{\citenamefont {Zhu}\ \emph {et~al.}(2008)\citenamefont {Zhu},
  \citenamefont {Zhang},\ and\ \citenamefont {Kaxiras}}]{ZHU_2008}%
  \BibitemOpen
  \bibfield  {author} {\bibinfo {author} {\bibfnamefont {W.}~\bibnamefont
  {Zhu}}, \bibinfo {author} {\bibfnamefont {Z.}~\bibnamefont {Zhang}}, \ and\
  \bibinfo {author} {\bibfnamefont {E.}~\bibnamefont {Kaxiras}},\ }\href
  {\doibase 10.1103/PhysRevLett.100.027205} {\bibfield  {journal} {\bibinfo
  {journal} {Physical Review Letters}\ }\textbf {\bibinfo {volume} {100}},\
  \bibinfo {eid} {027205} (\bibinfo {year} {2008})}\BibitemShut {NoStop}%
\bibitem [{\citenamefont {Chen}\ \emph {et~al.}(2009)\citenamefont {Chen},
  \citenamefont {Zhu}, \citenamefont {Kaxiras},\ and\ \citenamefont
  {Zhang}}]{ZHU_2009}%
  \BibitemOpen
  \bibfield  {author} {\bibinfo {author} {\bibfnamefont {H.}~\bibnamefont
  {Chen}}, \bibinfo {author} {\bibfnamefont {W.}~\bibnamefont {Zhu}}, \bibinfo
  {author} {\bibfnamefont {E.}~\bibnamefont {Kaxiras}}, \ and\ \bibinfo
  {author} {\bibfnamefont {Z.}~\bibnamefont {Zhang}},\ }\href {\doibase
  10.1103/PhysRevB.79.235202} {\bibfield  {journal} {\bibinfo  {journal}
  {Physical Review B (Condensed Matter and Materials Physics)}\ }\textbf
  {\bibinfo {volume} {79}},\ \bibinfo {eid} {235202} (\bibinfo {year}
  {2009})}\BibitemShut {NoStop}%
\end{thebibliography}


\end{document}